# Factors Controlling Oxygen Migration Barriers in Perovskites

Tam T. Mayeshiba
University of Wisconsin-Madison
1509 University Dr., Madison, Wisconsin, 53706

**Dane D. Morgan***
University of Wisconsin-Madison
1509 University Dr., Madison, Wisconsin, 53706
ddmorgan@wisc.edu

## Abstract

Perovskites with fast oxygen ion conduction can enable technologies like solid oxide fuel cells. One component of fast oxygen ion conduction is low oxygen migration barrier. Here we apply *ab initio* methods on over 40 perovskites to produce a database of oxygen migration barriers ranging from 0.2 to 1.6 eV. Mining the database revealed that systems with low barriers also have low metal-oxygen bond strength, as measured by oxygen vacancy formation energy and oxygen p-band center energy. These correlations provide a powerful descriptor for the development of new oxygen ion conductors and may explain the poor stability of some of the best oxygen conducting perovskites under reducing conditions. Other commonly-cited measures of space, volume, or structure ideality showed only weak correlation with migration barrier. The lowest migration barriers (< 0.5 eV) belong to perovskites with non-transition-metal B-site cations, and may require vacancy-creation strategies that involve no dopants or low-association dopants for optimal performance.

## 1. Introduction

Fast oxygen migration is a critical property for technologies that involve oxygen transport and exchange with the environment,[1] including solid oxide fuel cells (SOFCs), gas-separation membranes, oxygen sensors, chemical looping devices, and memristors. Many of the best oxygen ion conductors are perovskites, and searching this crystal class for fast, stable, and application-compatible oxygen conductors has been an active area of research for decades.[2, 3]

Optimized vacancy content and low oxygen migration barrier are typically the dominant factors leading to fast vacancy-mediated oxygen transport in perovskites. This paper assumes that doping strategies can optimize vacancy content[1, 4] and instead focuses solely on oxygen migration barrier. Sections 3 and 4 discuss briefly where information about vacancy concentration and dopant association effects would be valuable additional knowledge.

As an example of the scale of the effect of oxygen migration barrier on oxygen transport, for a hypothetical material at 1073K (800°C), a decrease of 0.4 eV in migration barrier energy corresponds to a 100 times increase in oxygen ion conductivity, potentially allowing a 100 times increase in the thickness of an SOFC electrolyte (see Supporting Information (SI), Section S2.1). Lowering the migration barrier by 0.6 eV at the same temperature could reduce the cathode area specific resistance (ASR) by 100 times (see SI, Section S2.2). These effects could enable significantly more stable, lower-temperature, cheaper SOFCs.

Descriptors for oxygen migration barrier could help identify low barrier materials prior to synthesis efforts. Previously-proposed descriptors for oxygen migration in perovskites include Kilner critical radius,[5-8] Goldschmidt tolerance factor,[5] volumetric factors,[5] crystal structure ideality,[9] oxygen vacancy formation energy,[10, 11] and metal-oxygen bond strength (fluorites [12]), both average[5] and related to vacancy trapping (summarized in Ref. [13]). This study focuses solely on perovskites, considers over 40 A-site and B-site combinations, uses oxygen migration barrier data from a single source and method, and evaluates dozens of descriptors at once, looking for a simple functional relationship between oxygen migration barrier and each proposed descriptor.

## 2. Methods

### 2.1. Choosing and evaluating descriptors

Because the descriptors are meant to be predictive, this study focuses on descriptors that do not require prior knowledge about the activated transition state. That is, we focus on descriptors from the bulk or initial defected state, and leave out descriptors which must be obtained from a transition-state calculation itself, such as transition state geometry[7] or actual path length.

Evaluating a descriptor consists of looking for the presence of a simple function, though not necessarily linear, between oxygen migration barrier and that descriptor. While combining descriptors might produce better correlations, evaluating descriptors individually provides a clear picture of the controlling physics without the risk of over-fitting.

Although fewer than two dozen descriptors appear in this study's main text and SI Section 3, we actually looked for correlation with hundreds of possible descriptors, of which only the most well-known, often-suggested, best-performing, and/or representative of key physics were included. These descriptors include those we have invented, those that have been proposed as important when comparing across crystal structures, and those that are commonly cited as being important within the perovskite crystal structure.[5, 9]

Most of the descriptors that were discarded were permutations of a single descriptor type, for example, the distance between a specific B-site cation and its neighboring oxygen in the positive z direction; in the negative z-direction; and so on.

## 2.2. Choosing systems

Compounds were generated and screened in the following order:
- $La^{3+}$[3d transition metal, excluding Cu and Zn, plus Ga]$^{3+}O_3$ due to the prevalence of 3d transition metals and Ga in known and studied devices.
- [Pr, Y]$^{3+}$[3d transition metal, excluding Cu and Zn, plus Ga]$^{3+}O^3$: the chemical similarity but different radii of $Pr^{3+}$ and $Y^{3+}$ compared to each other and to $La^{3+}$ allows A-site cation size effects to become evident.
- $La^{3+}$[Al, In, Tl]$^{3+}O_3$ due to the B-site similarity with fast conductors with B-sites Sc and Ga.

Assorted other compounds were partially or fully evaluated. Some were chosen because of A-site similarity to $La^{3+}$ and guidance from observed correlations that suggested low barrier materials, e.g., $SmGaO_3$. Others were chosen for 4d or 5d transition metal B-site cations similar to those previously studied, e.g. $LaRuO_3$. Finally, a few $A^{2+}B^{4+}O_3$ systems were evaluated to check the behavior of the correlations in the $A^{2+}B^{4+}O_3$ system.

## 2.2. Computational Methods

Vacancy-mediated oxygen diffusion in these systems was modeled by a single hop from oxygen position o29 to o30, which should be taken to be sampling possible hopping barriers that range over 0.4 eV based on calculations of all hops within three supercells (see Ref.[14] electronic supplementary information (ESI) Figure S2.1 for atomic positions, and Ref. [14] ESI Section S8 and this work's SI Section S4 for the range of hops). The exception was the barrier for $LaRuO_3$, where we believe specific geometry in that hop forced it to exceed the range of reasonable migration barrier values for the system as a whole; therefore, the o29 to o30 migration barrier (1.847 eV) was substituted out and an in-plane migration barrier (oxygen position o31 to o30) was used instead (1.428 eV).

Climbing nudged-elastic-band calculations[15, 16] were automated using the MAterials Simulation Toolkit (MAST)[17, 18] and performed using the Vienna Ab-initio Simulation Package (VASP) [19-23] on a 2x2x2 formula unit supercell where internal relaxation allowed octahedral tilting,[24, 25] and with settings and approaches, including pseudopotentials picked for accuracy, as described previously.[14]

GGA+U[26-29] calculations were not used in this study due to the uncertainty in selecting U values,[30-32] the computational expense of performing such selection for a wide variety of cations, and convergence issues in GGA+U calculations, particularly in Co and Ni. Due to the cancellation of errors between initial and activated states in migration barrier calculations, and due to the absence of any redox occurring in our models, we expect GGA calculations to be fairly accurate for the key values in this paper. The decision to use GGA methods has also been made in other perovskite migration barrier studies.[7, 11].

Because mobile oxygen vacancies in a typical host perovskite will be charge-compensated.[1, 33, 34], all migration barriers discussed in the main text use a charge-compensated oxygen vacancy, where the electrons donated by the vacancy are removed from the system either through explicit electron-removal or through explicit Strontium doping on the A-site, as described previously.[14] (See the SI for more discussion.) Table A.1 also includes non-charge-compensated data, in which the electrons donated from the vacancy are left in the system, often to reduce a nearby B-site cation.

Vacancy formation energies were calculated as in Lee et al.[35] Where the oxygen vacancy was charge-compensated by explicit electron removal, an extra term was applied to the vacancy formation energy in order to account for the removal of those electrons to the electron reservoir of the bulk, as in Section 3.1 of Lin et al.[36] The potential alignment correction in such cases used, for each species, the mean for all atoms of that species in the supercell. These vacancy formation energies are intended for use as descriptors rather than as energy values that can be compared to experiments, and they may not be accurate for the latter. In particular, no finite-size scaling correction was applied, the errors in the GGA approximations may affect the absolute value of each vacancy formation energy,[35] and the given oxygen partial pressure and temperature, while possible in an SOFC, may not be appropriate for all systems. Furthermore, the values do not take into account the full defect chemistry model of the perovskites and therefore are in most cases not representative of experimental vacancy formation energies. For the purposes of this study, the important value is the relative difference in oxygen vacancy formation energy between different systems.

Radii for use in descriptors were picked from the Shannon crystal radii[11, 37] based on an estimate of high-spin and low-spin state from undefected bulk magnetic moment calculations, using 6-coordination for $O^{2-}$, $B^{3+}$, and $B^{4+}$, and 12-coordination, or 9-coordination if 12-coordination was not available, for $A^{2+}$ and $A^{3+}$. For doped systems, the A-site radius was the weighted average of the radii for the A-site occupations.

## 3. Results

This section compares the calculated barriers with experiment where applicable, then plots migration barrier versus Mendeleev number to show the scale of the results, and then plots migration barrier versus the two best-performing descriptors, oxygen vacancy formation energy and oxygen p-band center energy. SI Section 3 contains the results for the other descriptors.

Each oxygen hop followed a curved migration path around a B-site cation.[13, 38] Calculating all oxygen hops for three systems indicates that each hop should be considered to have been sampled from an approximately uniformly distributed range of about 0.4 eV (see Ref. [14] ESI Section S8, and SI for this paper, Figure S4.1 and Figure S4.2). Dopant positions may also contribute to a spread of barriers, with a spread estimated by select studies to be about 0.3 eV (see Ref.[14], ESI Section S7). Supercell size is estimated to have a non-systematic effect of approximately 0.1 eV (see Table A.1).

Figure 1 compares our calculations to migration barriers from experimental data, whose details are given in SI Table S1.1. (For a comparison with calculated rather than experimental literature data, see SI Figure S1.2.) For each computed value there is a comparable experimental value within the range bar of the calculation. However, the large spread in the experimental values and the uncertainty in the calculations makes any quantitative comparison very difficult. We note that many of the non-transition-metal B-site cations have predicted values lower than any of the experiments, and some of our predictions for this class of compounds (discussed in Section 4.3) also have very low values. Although these values are within the range bars of the experiments, they may also be lower because of dopant effects. Where the calculations were performed on undoped systems, with the vacancy compensated as though it had been created with dopants, which would be a requirement for these non-transition-metal B-site systems,[3, 4] the lack of any dopant association terms may lead to lower predicted migration barriers than in experiment.[6, 39, 40]

For LaGaO$_3$, the calculation was also performed on a doped system but was still low compared to experiment. In this case, the range bar for doped hops may be necessary to represent all of the hops, including higher barrier ones, which might be necessary barriers to overcome to follow a percolating pathway in a real system. Also, the relatively high dopant concentration (x=0.25) and relatively small calculation supercell size mean that in periodic space, the calculation has actually formed an ordered structure, which may not be comparable to a disordered structure with lower dopant concentration (x=0.1 for LSGM[4]).

However, there is good agreement between our calculations and experiment for LaScO$_3$, LSS, LSSM, and BaTiO$_3$, which also fall under the same category of non-transition-metal B-site cation. These results suggest that a detailed system-by-system comparison is necessary to identify all of the sources of differences between the systems and experiment, possibly through an exploration of different calculation parameters, for example, pseudopotential choice, supercell size, and exact doping concentration.

In summary, the primary sources of differences between the calculations and experiments may be: (1) system-specific, for example relating to the calculation parameters of a given species, and which may require in-depth system-specific studies, and (2) model-based, for example, the absence of dopants and dopant-vacancy interactions in most of the calculated systems; the high-concentration of dopants and periodic ordering of dopants in the nine calculated doped systems; the non-dilute concentration of vacancies in most of the calculated systems (1/24 oxygen sites in a 2x2x2 size supercell), which would be related in experiments to the dopant concentration, B-site cation species, temperature, and oxygen partial pressure; and the use of a single hop which does not capture full information on the percolating pathway for diffusion, where long-range diffusion is controlled by the rates of single atomic hops but is the result of many such hops in series.[41, 42] Interrelations among all of these differences make it difficult to make an overall statement on our calculated migration barriers versus the experimental migration barriers except to say that in general, we would not expect a single hop to be directly comparable to the migration barrier generated from experimental data, so the range bars provided should be used when comparing to experimental data.

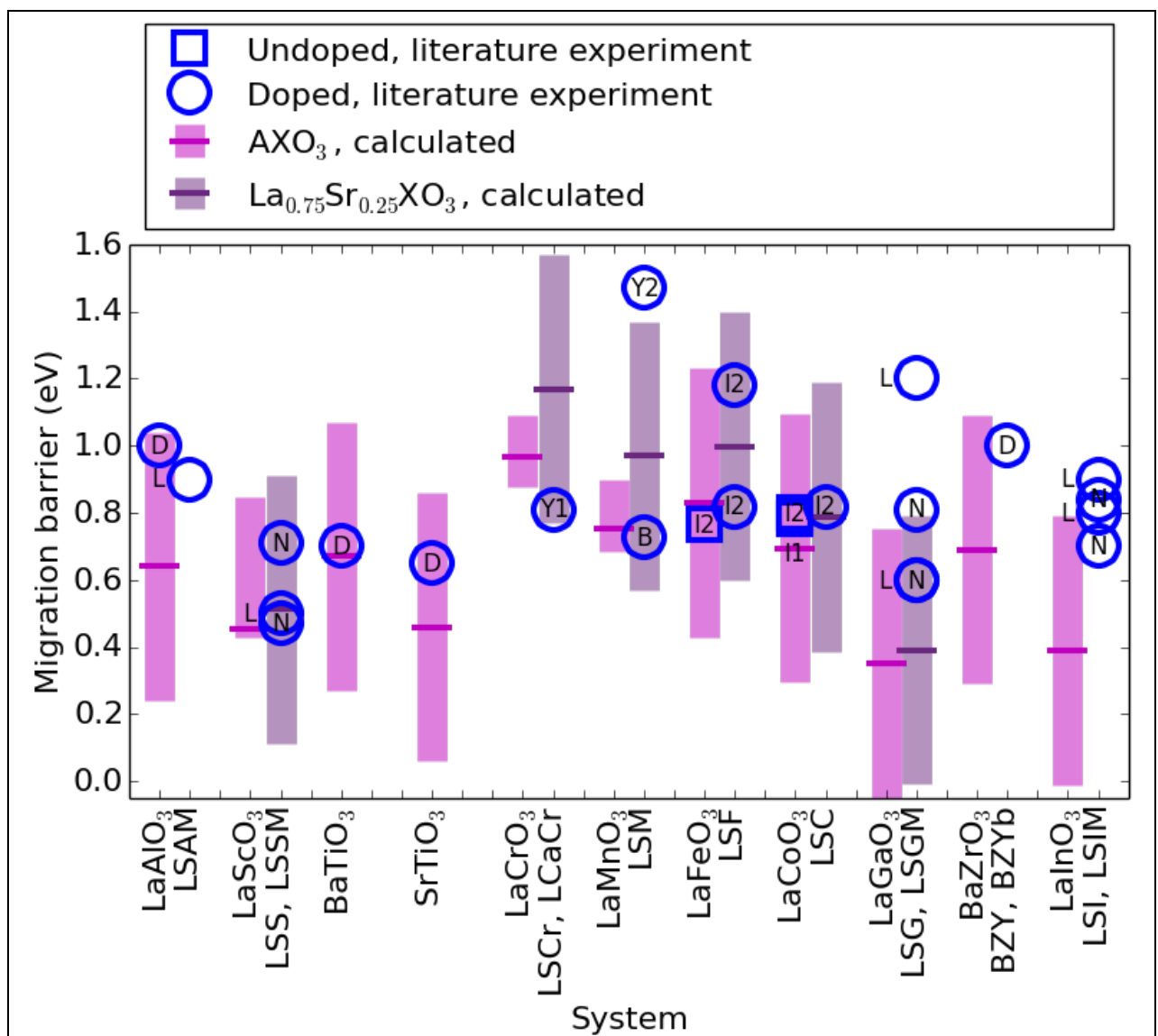

Figure 1. Literature oxygen migration barriers compared with calculated oxygen migration barriers. Literature value references are abbreviated by the first initials of the author's last name: B=[43], D=[44] (for this reference when a comparison is made to an undoped system calculation the experiment contains weak acceptor doping), I1=[45], I2=[34], L=[3], N=[33], Y1=[46], Y2=[47]. Details on the literature values are given in the Supporting Information, Table S1.1. For $AXO_3$ calculated data, range bars are given for 96 hops in B-site cations Sc, Cr, and Mn, with the common hop shown at the correct position in the range, and are estimated at +/- 0.4 eV for the remaining B-site cations. For $La_{0.75}Sr_{0.25}XO_3$ calculated data, the range over all hops is estimated at +/- 0.4 eV. For doped system labels, "S" refers to Sr and a fourth letter "M" refers to Mg, for systems $A_{1-x}A'_xB_{1-y}B'_y$ with $0.1 \leq x \leq 0.3$ and $0 \leq y \leq 0.1$.

Figure 2 presents our migration barrier data plotted against B-site cation Mendeleev number (see Figure 1 in Ref. [48]). (See SI Figure S3.A1 for the data plotted against atomic number, and SI Figure S3.A2 for the data plotted against a more spread-out version of Mendeleev number.) Migration barriers generally fall moving through the B-site cation Mendeleev numbers 50-81 with particularly low migration barriers calculated for non-transition-metal B-site cations (Sc, Ga, Al, In, Tl, Y). More discussion on extremely low barriers will be given in Section 4, Discussion.

The calculated migration barrier values cover a wide range of over 1.4 eV, and include some particularly low values. To understand the origin of this spread and these low values, we looked for correlation with hundreds of possible descriptors, focusing on over 20 of the most well-known or representative of key physics, as described in the Methods section. We find weak or no correlation between oxygen migration barrier and descriptors relating to space, volume, or ideality of the structure (see SI Section 3). However, we find strong correlation between oxygen migration barrier and descriptors related to metal-oxygen bond strength. A correlation of oxygen transport with metal-oxygen bond strength has been proposed several times in the literature[9, 49-51] and our large database now allows these relationships to be definitively identified. Here we focus on two measures of metal-oxygen bond strength, the vacancy formation energy and oxygen p-band center energy.

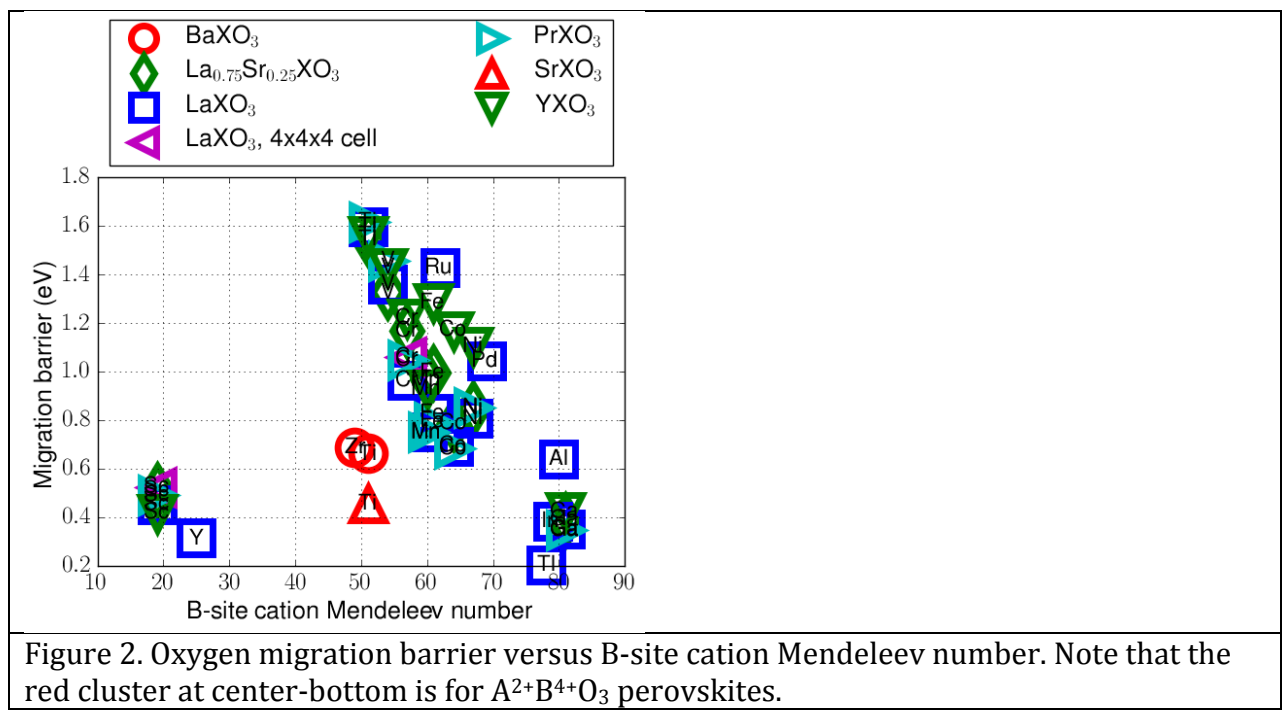

Figure 2. Oxygen migration barrier versus B-site cation Mendeleev number. Note that the red cluster at center-bottom is for $A^{2+}B^{4+}O_3$ perovskites.

Figure 3 shows that oxygen vacancy formation energy correlates strongly with oxygen migration barrier. While oxygen vacancy formation energy has been proposed as a descriptor for oxygen diffusion before,[10, 11, 52] our data shows that the correlation holds over dozens of perovskite systems. Kwon et al. studied oxygen vacancy formation energy over a subset of B-site cations with alloying and did not note the correlation between oxygen vacancy formation and oxygen migration barrier. However, an examination of their x=0 (undoped) values shows the same trends as in Figure 3 (blue squares, Cr, Mn, Fe, Co, Ni), although the correlation we observe would not be apparent from just those values.[52] The root-mean-square (RMS) error between calculated migration barrier and migration barrier as a result of a linear fit with oxygen vacancy formation energy is just 0.2 eV, and also has an average of 0.2 eV cross-validation score (with a standard deviation of 0.02 eV) over 1000 2-fold (leave-out-half the data, chosen randomly, for testing) cross validation tests.[53] The quite small cross-validation score and its closeness to the RMS error of the fit to all of the data strongly support the validity of the descriptor.

The observed correlation between migration and vacancy formation energy can be understood physically by noting that lower oxygen vacancy formation energy indicates that it is easier to break cation bonds with oxygen, which is a necessary part of the process for migrating an oxygen atom from one position to another. This correlation is conceptually useful, and of some practical value for computations as vacancy formation energies are somewhat faster to determine than migration barrier.

These oxygen vacancy formation energies are not comparable to experimentally observed oxygen vacancy formation energies, as the method used is chosen to faithfully represent the energy of breaking bonds to create the vacancy, not the often complex defect coupling that controls observed experimental vacancy formation energies. The values should therefore be taken as relative values (see Section 2.2, Computational Methods). Furthermore, the values have a possible spread of some +/- 0.4 eV due to the range of values introduced by symmetry-distinct oxygen sites and varied dopant positions (see SI Section S4).

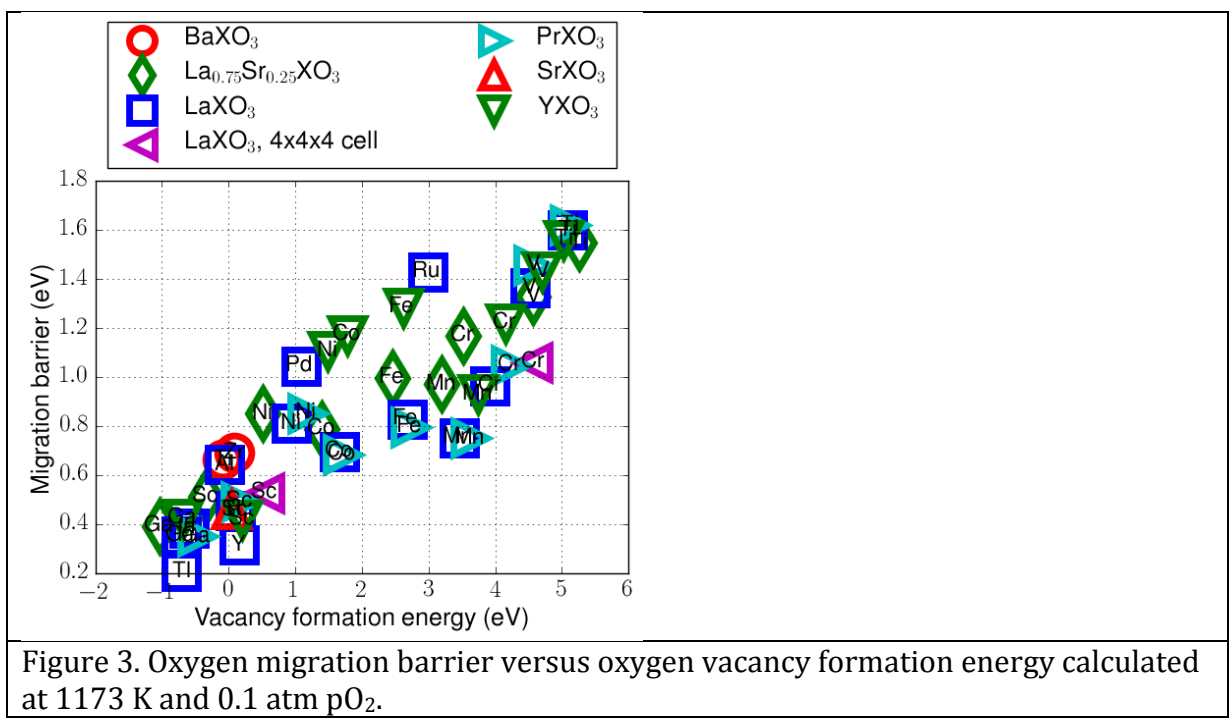

Figure 3. Oxygen migration barrier versus oxygen vacancy formation energy calculated at 1173 K and 0.1 atm $pO_2$.

Figure 4 shows the oxygen migration energy as a function of the oxygen 2p-band center position relative to the Fermi level, which we abbreviate as oxygen p-band center energy. This quantity has recently been shown to correlate with many important oxygen kinetics properties in perovskites, including calculated oxygen defect and adsorption energetics[54] and experimentally measured oxygen surface exchange rates,[54] area specific resistance,[54] activation energies of surface exchange and diffusion,[55] and work functions.[56] As pointed out in Ref. [54], the oxygen p-band is also a measure of metal-oxygen bond strength, which correlation likely emerges because increasing p-band center energy correlates with less charge transfer from cations to oxygen and weaker bonding between them. Therefore, the strong negative correlation shown in Figure 4 further supports the hypothesis that the oxygen migration is largely controlled by the strength of the oxygen-cation bonds.

The RMS error between calculated migration barrier and migration barrier as a result of a linear fit with oxygen p-band center is just 0.2 eV, and also has an average of 0.2 eV cross-validation score (with a standard deviation of 0.02 eV) over 1000 2-fold (leave-out-half the data, chosen randomly, for testing) cross validation tests.[53] The quite small cross-validation score and its closeness to the RMS error of the fit to all of the data strongly support the validity of the descriptor.

Note, however, that neither correlation with oxygen vacancy formation energy or p-band center energy is particularly strong for moderate-barrier materials: a vacancy formation energy of about 3 eV or an oxygen p-band center energy of about -4.5 eV span approximately 0.7 eV of migration barriers. Therefore, these correlations may be useful more as guides than as quantitative predictors.

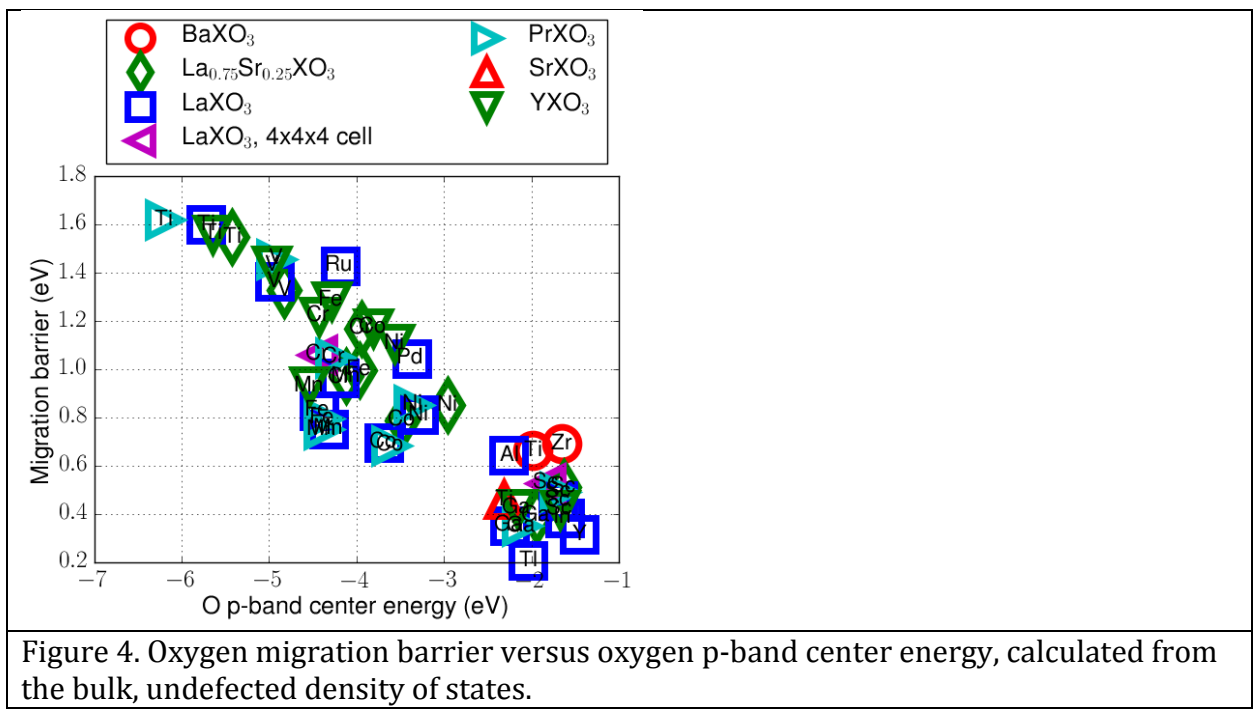

Figure 4. Oxygen migration barrier versus oxygen p-band center energy, calculated from the bulk, undefected density of states.

## 4. Discussion

### 4.1. Oxygen p-band center

The oxygen p-band center is a particularly useful descriptor as it can be rapidly calculated for a new system, making it a valuable tool for rapid screening. As the oxygen p-band correlates well with both vacancy formation energy (see SI Figure S1.1 and Refs.[54, 57]) and migration barrier (Figure 4), the oxygen p-band center energy should also correlate well with the activation energy for diffusion. This correlation has recently been shown in Ref. [55]. We therefore expect that the oxygen p-band center energy correlates well with the overall diffusion coefficient $D$, although the uncertainty in overall diffusion value measurements has so far made this correlation hard to establish, as discussed in Ref. [55]. Furthermore, Lee et al. show that oxygen p-band center energy correlates with the surface exchange coefficient, $k^*$.[54] If the oxygen p-band center energy correlates with $D$ and $k^*$, then from the Adler-Lane-Steele model (see SI Section S2.2),[58] the oxygen p-band center energy should correlate with the overall cathode area-specific resistance (ASR), also consistent with what is shown in Lee et al.[54] The similar success of these many coupled correlations demonstrates the strong consistency in these correlative relationships, shows the power of this descriptor for modeling SOFCs and general active-oxygen materials, and supports the assertion that metal-oxygen bond strength plays a dominant role in the performance of these materials.

An important implication of the above results is that the fast oxygen ion conductors will typically have the weakest cation-oxygen bonds, which will make them particularly susceptible to reduction. Thus this correlation can explain the tendency of many fast oxygen ion conductors to have problems with second phase formation and increased electronic conductivity from defects when used as electrolytes exposed to reducing anodic conditions.[59]

### 4.2. Defining descriptors within a scope

We were initially surprised by the non-correlation of geometric factors over our dataset, given the emphasis on geometry in the literature. On further reflection, the finding makes sense considering, on the one hand, our large variety of A-site and B-site cations, and on the other hand, our relatively narrow constraints of using mostly $A^{3+}B^{3+}O_3$ perovskites that, when doped, had only a single doping level, single dopant type, and fixed dopant positioning.

Some geometric descriptors may be more effective in a cross-class materials search, where there are dramatic changes across structure classes. For example, Sammells et al. found a V-shaped correlation between activation energy and free volume, with a negative correlation for $A^{2+}B^{4+}O_3$ perovskites and a positive correlation for fluorites[5]; our database did not have free volumes as high as those for the fluorites. Our study also looked only at single crystal diffusion, while polycrystalline diffusion and grain boundary diffusion might introduce additional geometric and volumetric factors.

Other geometric descriptors may be more effective in a very narrow subset of systems. For example, Mogensen et al. considered structure stress to be of more importance than bond strength, but their study only spanned perovskites with non-transition-metal B-site cations in Group 3A (Sc, Ga, Al, In).[60] Similarly, SI Section S3 shows many separations by A-site cation, so looking over several A-site cations at a single B-site cation may well show trends like in Ranløv et al. (reproduced in Ref. [9] Figure 4). Structure stress, crystal structure ideality, and volume descriptors may work best when looking at fine changes in an otherwise very similar geometry.

This understanding that descriptors may apply only within a a certain scope of materials helps unify our understanding of descriptors in perovskites by suggesting that there may be several best or best combinations of descriptors, each most appropriate to a specific search space and the application context. For the combinatorial problem of a single structure class with many compositions, which we undertook here, geometry may be similar enough that large changes do not dominate, while not so similar that very fine changes can be differentiated, and therefore geometric descriptors show only weak correlation. The descriptors of metal-oxygen bond strength proved the most useful for extrapolation of oxygen ionic migration barrier within the perovskite structure class, across many compositions.

### 4.3. Low-barrier materials

Several of our calculated oxygen migration barriers for non-transition-metal B-site cations are significantly below 0.5 eV (e.g. 0.21 eV, $LaTlO_3$, and 0.32 eV, $LaYO_3$, see Table A.1), Low barriers are expected in the region of non-transition-metal B-site cations,[3] but no barriers this low have been reported from previous experimental studies.

Dopants may be the main reason for the discrepancy between the lower bound of the calculated migration barriers and the lower bound of migration barriers seen in experiment. While these systems lack dopants in the calculation, they are precisely the systems that usually require aliovalent dopants to enable oxygen migration. The extremely low calculated migration barriers might then be regarded as an ideal lower-bound barrier for oxygen conduction, in a system where the moving vacancy is compensated as though it were created by aliovalent dopants, but is not hampered by dopant-vacancy interactions.

Significant previous work has been done to reduce dopant-vacancy association energies, notably through the host-dopant size studies.[6, 9] One strategy to eliminate dopant-vacancy association might be to form vacancies in nearby layers of doped or undoped transition metal B-site perovskites. Figure 3 shows that non-transition-metal B-site perovskites have the lowest compensated vacancy formation energies, so oxygen vacancies may collect in these materials from neighboring materials in which the vacancies were created. Epitaxial strain could also be used to induce preferential vacancy segregation into the low-barrier material.[61] These and other strategies, combined with advanced materials growth and deposition technologies could yield exceptional fast oxygen ion conductors in this class of non-transition-metal B-site cation perovskites.

## 5. Conclusions

We find that a database of $A^{3+}B^{3+}O_3$ (and select $A^{2+}B^{4+}O_3$) perovskites relevant for oxygen ion transport and exchange applications have calculated oxygen migration barriers ranging from 0.2 to 1.6 eV, with the lowest barrier values calculated for non-transition-metal 3+ cations on the B-site (Sc, Al, Ga, In, Tl). Notably, low barriers were predicted from our trends and found for $LaTlO_3$ and $(Sm, Dy, Er)(Al, Ga)O_3$. Such low barriers may only be accessible without significant dopant-vacancy interaction and their use as an estimate of the overall oxygen vacancy diffusion barrier is subject to significant uncertainty. Although these systems are also likely to require doping in order to produce a significant amount of vacancies, the low barriers may be accessible if such materials could be created without vacancies. The magnitude of the range and the physics of the migration appear to be dominated by metal-oxygen bond strength, as measured by oxygen vacancy formation and oxygen p-band center energy. Other descriptors relating to space, volume, and ideality of the structure do not correlate well with oxygen migration barrier in this dataset.

**Acknowledgments**
The National Science Foundation (NSF) Graduate Fellowship Program under Grant No. DGE-0718123 and the UW-Madison Graduate Engineering Research Scholars Program provided funding for T. Mayeshiba. Calculations utilized computing resources of the Extreme Science and Engineering Discovery Environment (XSEDE), which is supported by NSF grant number OCI-1053575, and from the UW-Madison Center for High Throughput Computing (CHTC) in the Department of Computer Sciences. The CHTC is supported by UW-Madison, the Advanced Computing Initiative, the Wisconsin Alumni Research Foundation, the Wisconsin Institutes for Discovery, and the NSF, and is an active member of the Open Science Grid, which is supported by the NSF and the U.S. Department of Energy's Office of Science. Support for D. Morgan and for the MAST tools applied in this work were provided by the NSF Software Infrastructure for Sustained Innovation (SI$^2$) award No. 1148011. T. Mayeshiba would like to thank Dr. Henry Wu for fruitful discussion, and Dr. Glen Jenness for sharing his oxygen p-band center energy code.

**Supporting Information**. Notes for literature values, figure for oxygen p-band energy versus oxygen vacancy formation energy, derivation of the effect of oxygen migration barrier on oxygen conductivity and cathodic ASR, and figures for additional descriptors, including Goldschmidt tolerance factor, Kilner critical radius, percent free volume, cohesive energy, Fermi energy, and bandgap, among many others.

# Appendix

Table A.1. Complete table of oxygen migration barrier data calculated for perovskite systems, including data for charge-compensated and non-charge compensated vacancies. See Section 2, Computational Methods and SI Section S1 for an explanation of compensation. Italics denote an $A^{2+}B^{4+}O_3$ perovskite; all others are $A^{3+}B^{3+}O_3$ or $A^{3+}_{0.75}A'^{2+}_{0.25}B^{3+}O_3$. Systems evaluated for low migration barrier based on correlations may be missing some data, as they were not part of the original calculation set.

| System | Oxygen migration barrier (eV) | | Oxygen 2p-band center energy (eV) | | Oxygen vacancy formation energy (eV) at 1173K, 0.1 atm $p(O_2)$ | |
|---|---|---|---|---|---|---|
| | Charge-compensated | Non-charge-compensated | Charge-compensated | Non-charge-compensated | Charge-compensated | Non-charge-compensated |
| *BaTiO$_3$* | 0.67 | | -1.99 | | -0.09 | |
| *BaZrO$_3$* | 0.69 | | -1.66 | | 0.08 | |
| DyAlO$_3$ | 0.55 | | -2.30 | | 0.04 | |
| DyGaO$_3$ | 0.44 | | -2.09 | | -0.65 | |
| ErAlO$_3$ | 0.59 | | -2.33 | | 0.11 | |
| ErCoO$_3$ | 1.22 | | -3.98 | | 1.08 | |
| ErGaO$_3$ | 0.47 | | -2.05 | | -0.62 | |
| La$_{0.75}$Sr$_{0.25}$CoO$_3$ | 0.79 | | -3.48 | | 1.40 | |
| La$_{0.75}$Sr$_{0.25}$CrO$_3$ | 1.17 | | -3.95 | | 3.52 | |
| La$_{0.75}$Sr$_{0.25}$FeO$_3$ | 1.00 | | -3.96 | | 2.46 | |

| Compound | | | | | | |
|---|---|---|---|---|---|---|
| La$_{0.75}$Sr$_{0.25}$GaO$_3$ | 0.39 | | -1.94 | | -1.04 | |
| La$_{0.75}$Sr$_{0.25}$MnO$_3$ | 0.97 | | -4.12 | | 3.20 | |
| La$_{0.75}$Sr$_{0.25}$NiO$_3$ | 0.85 | | -2.96 | | 0.51 | |
| La$_{0.75}$Sr$_{0.25}$ScO$_3$ | 0.51 | | -1.64 | | -0.34 | |
| La$_{0.75}$Sr$_{0.25}$TiO$_3$ | 1.55 | | -5.42 | | 5.26 | |
| La$_{0.75}$Sr$_{0.25}$VO$_3$ | 1.33 | | -4.83 | | 4.57 | |
| LaAlO$_3$ | 0.64 | | -2.26 | | -0.06 | |
| LaCoO$_3$ | 0.70 | 0.76 | -3.69 | -3.73 | 1.66 | 1.97 |
| LaCrO$_3$ | 0.97 | 1.67 | -4.19 | -4.24 | 3.92 | 4.29 |
| LaCrO$_3$, 4x4x4 supercell | 1.06 | 1.66 | -4.44 | -4.44 | 4.57 | 4.29 |
| LaFeO$_3$ | 0.83 | 0.81 | -4.43 | -4.49 | 2.67 | 2.66 |
| LaGaO$_3$ | 0.35 | 2.07 | -2.25 | -1.95 | -0.71 | 4.19 |
| LaInO$_3$ | 0.39 | | -1.63 | | -0.60 | |
| LaMnO$_3$ | 0.75 | 0.94 | -4.32 | -4.42 | 3.46 | 3.43 |
| LaNiO$_3$ | 0.81 | 0.90 | -3.29 | -3.29 | 0.94 | 1.22 |
| LaPdO$_3$ | 1.04 | | -3.37 | | 1.08 | |
| LaRuO$_3$ | 1.43 | | -4.18 | | 2.99 | |
| LaScO$_3$ | 0.46 | 1.97 | -1.70 | -1.70 | 0.10 | 5.59 |
| LaScO$_3$, 4x4x4 supercell | 0.53 | 1.86 | -1.84 | -1.84 | 0.54 | 5.58 |
| LaTiO$_3$ | 1.60 | 1.61 | -5.73 | -5.76 | 5.09 | 5.05 |
| LaTlO$_3$ | 0.21 | | -2.04 | | -0.71 | |
| LaVO$_3$ | 1.36 | 1.63 | -4.94 | -5.09 | 4.52 | 4.62 |
| LaYO$_3$ | 0.32 | | -1.45 | | 0.15 | |
| PrCoO$_3$ | 0.68 | 0.76 | -3.61 | -3.61 | 1.71 | 2.04 |
| PrCrO$_3$ | 1.05 | 2.00 | -4.25 | -4.25 | 4.23 | 4.33 |
| PrFeO$_3$ | 0.80 | 0.83 | -4.38 | -4.38 | 2.73 | 2.70 |
| PrGaO$_3$ | 0.35 | 2.13 | -2.11 | -2.11 | -0.48 | 4.26 |
| PrMnO$_3$ | 0.75 | 1.03 | -4.39 | -4.39 | 3.63 | 3.48 |
| PrNiO$_3$ | 0.85 | 0.88 | -3.35 | -3.35 | 1.17 | 1.40 |
| PrScO$_3$ | 0.49 | 2.03 | -1.69 | -1.69 | 0.17 | 5.51 |
| PrTiO$_3$ | 1.62 | 1.54 | -6.20 | -6.20 | 5.11 | 4.86 |
| PrVO$_3$ | 1.46 | 1.75 | -4.92 | -4.92 | 4.57 | 4.62 |
| SmAlO$_3$ | 0.55 | | -2.25 | | -0.13 | |
| SmCuO$_3$ | 0.66 | | -2.99 | | 0.23 | |
| SmGaO$_3$ | 0.36 | | -2.00 | | -0.56 | |
| *SrTiO$_3$* | 0.46 | | -2.32 | | 0.05 | |
| YCoO$_3$ | 1.17 | 1.47 | -3.81 | -3.81 | 1.78 | 1.92 |
| YCrO$_3$ | 1.22 | 2.16 | -4.43 | -4.43 | 4.16 | 4.45 |
| YFeO$_3$ | 1.29 | 0.97 | -4.28 | -4.28 | 2.62 | 2.65 |
| YGaO$_3$ | 0.43 | 2.33 | -2.16 | -2.16 | -0.68 | 4.34 |
| YMnO$_3$ | 0.93 | 1.26 | -4.54 | -4.54 | 3.74 | 3.51 |
| YNiO$_3$ | 1.11 | 1.22 | -3.57 | -3.57 | 1.49 | 1.55 |
| YScO$_3$ | 0.42 | 1.90 | -1.68 | -1.68 | 0.20 | 5.59 |
| YTiO$_3$ | 1.56 | 1.57 | -5.64 | -5.64 | 5.02 | 5.13 |
| YVO$_3$ | 1.43 | 1.90 | -4.97 | -4.97 | 4.70 | 4.74 |

SUPPORTING INFORMATION
for
**Factors Controlling Oxygen Migration Barriers in Perovskites**

Tam Mayeshiba, Dane Morgan
April 21, 2016

**Table of Contents**

## S0. Introduction

This supporting information contains accompanying derivations, tables, and figures to the main text.

## S1. General supporting information

Tables and Figures designated with S1 denote general supporting information referenced from the main text.

### S1.1 Charge compensation (vacancy charge)

For an $A^{3+}B^{3+}O_3$ aliovalently-doped perovskite, vacancies are expected to be created through the doping process[1], resulting in $V_O^{\bullet\bullet}$ that are charge-compensated by the aliovalent dopants.

We model this situation first by using aliovalently-doped perovskites. Here, the $La_{0.75}Sr_{0.25}XO_3$ perovskites contain two Sr atom substitutions on the 8 A sites in the supercell. For the undefected supercell, these substitutions nominally oxidize nearby B-site cations to 4+. When the oxygen vacancy defect is created, simply by removing an oxygen atom, the two electrons donated by the oxygen nominally reduce those B-site cations from 4+ back to 3+.

The concentration of oxygen vacancies in experimental systems is often not x/2, where x is the formula-unit Strontium concentration.[2] The doped supercells in our calculations are intended to model the charge-compensation of the oxygen vacancy, where two $A^{2+}$ are required to charge compensate a single vacancy, rather than to make a statement on the oxygen non-stoichiometry behavior of their particular perovskites.

Another way to model a charge-compensated oxygen vacancy is to not include any explicit aliovalent dopants but to remove those two electrons donated by the oxygen along with the oxygen atom. This action is analogous to either removing those two electrons to oxidized B-site cations somewhere outside the supercell, or simultaneously introducing aliovalent A-site dopants somewhere outside the supercell immediately upon creating the vacancy. This method requires a +2 charged supercell calculation, and we previously showed in Ref.[3] Electronic Supporting Information (ESI) Section S7 that explicitly doped supercells and charge-compensated supercells generate migration barriers that are similar to within 0.25 eV, which is also within the total range of symmetry-distinct migration barriers for a system (Section S4, Ref.[3] ESI Section S8).

The migration of a non-charge-compensated vacancy is probably the less realistic scenario for an SOFC. For the non-charge-compensated vacancy data, the effect of the vacancy is nominally to reduce neighboring B-site cations from 3+ to 2+.

Table A.1 in the main paper shows that the difference between charge-compensated and non-charge compensated oxygen migration barriers is within 300 meV of each other for B-site cations easily able to adopt a 2+ oxidation state: Mn, Ni, Fe, Co. On the other hand, the difference between charge-compensated and non-charge-compensated oxygen migration barriers is very large, over 1.5 eV, for B-site cations that only exist in a 3+ oxidation state: Sc, Ga. In the latter case, the non-charge-compensated oxygen vacancy is probably not physical.

### S1.2. Supercell size

Migration barrier changed 0.1 eV or less with supercell size for both charge-compensated and non-charge-compensated vacancies. Oxygen migration seems to be dominated by local effects (local distortions and local charge patterns). Also, supercell energies for the transition state were always compared with supercell energies for the initial state for a supercell of the same size and charge state, so systematic finite-size effects in energy of charged and uncharged supercells may have cancelled out.

## S2. Derivations for the effect of a change in migration barrier on SOFC metrics

In the following subsections, we derive the effects of changes in oxygen migration barrier on the SOFC metrics of oxygen ionic conductivity (Section S2.1) and cathodic area-specific resistance (ASR, Section S2.2).

Oxygen migration barrier, which we take equivalently as vacancy migration barrier, can be given either as an experimental enthalpy or as a calculated energy. See the Electronic Supporting Information (ESI) from Ref. [3], Section S9, for the applicability of comparing $H_{mig}(P_{constant})$ given in the following equations with calculated $E_{mig}(V_{constant})$.

To isolate the effects of migration barrier alone, we assume two perovskite systems where oxygen vacancy concentration was tuned separately using dopants, and where tortuosity, porosity, surface area, lattice parameter, number of oxygen sites per unit volume, and correlation factor are similar.

Note that lowering migration barrier will increase conductivity and decrease ASR.

### S2.1. Effect of change in migration barrier on oxygen conductivity

In this section, we estimate the effect of a change in oxygen migration barrier on oxygen ion conductivity.

Oxygen ionic conductivity for perovskites with vacancy-mediated diffusion can be given by Equation S2.1.1, using the Nernst-Einstein equation,[1] and an Arrhenius expression for the diffusion coefficient, where $\sigma_i$ is the ionic conductivity, $\mu_v$ is the vacancy mobility, $C_v$ is the vacancy concentration, $q_v$ is the charge on the vacancy, $D_v$ is the vacancy diffusion coefficient, $A_{0,v}$ is the geometric and vibrational prefactor for the vacancy diffusion coefficient, and $H_{mig}$ is the vacancy migration enthalpy. For a more complete derivation, see the ESI to Ref. [3].

$$\sigma_i = C_v q_v \mu_v = \frac{C_v q_v^2 D_v}{kT} = \frac{C_v q_v^2 A_{0,v} \exp\left(\frac{-H_{mig}}{kT}\right)}{kT} \quad \text{Equation S2.1.1}$$

Using Equation S2.1.1 and the assumptions of similarity given in the introduction to Section S2, that is, equal prefactor terms and vacancy concentrations, Equation S2.1.2 can be used to give the ratio of two ionic conductivities in terms of migration energies. Specifically, Equation S2.1.2 simplifies to Equation S2.1.3. Taking the natural logarithm of both sides gives Equation S2.1.4.

$$\frac{\sigma_{i,1}}{\sigma_{i,2}} = \frac{C_{v,1} q_v^2 A_{0,v,1} \exp\left(\frac{-H_{mig,1}}{kT}\right)}{C_{v,2} q_v^2 A_{0,v,2} \exp\left(\frac{-H_{mig,2}}{kT}\right)} \approx \frac{\exp\left(\frac{-H_{mig,1}}{kT}\right)}{\exp\left(\frac{-H_{mig,2}}{kT}\right)} \quad \text{Equation S2.1.2}$$

$$\frac{\sigma_{i,1}}{\sigma_{i,2}} \approx \exp\left(\frac{-H_{mig,1}}{kT} + \frac{H_{mig,2}}{kT}\right) = \exp\left(\frac{H_{mig,2}}{kT} - \frac{H_{mig,1}}{kT}\right) \quad \text{Equation S2.1.3}$$

$$\ln\left(\frac{\sigma_{i,1}}{\sigma_{i,2}}\right) \approx \frac{H_{mig,2} - H_{mig,1}}{kT} \quad \text{Equation S2.1.4}$$

This relationships can be conveniently written in terms of orders of magnitude as follows. In Equation S2.1.5 we define $M$ by the relationship that the ionic conductivity of the first system is $M$ orders of magnitude greater than the ionic conductivity of the second system. Substituting Equation S2.1.5 into Equation S2.1.4 gives Equation S2.1.6. Converting the natural logarithm to a base-10 logarithm gives Equation S2.1.7. From Equation S2.1.7, we can see that if the ionic conductivity of the first system at 1073K, or $kT \approx 0.0925 \, eV$, is two orders of magnitude greater than that of the second system, or $M = 2$, then that difference corresponds to a migration barrier about 0.43 eV smaller for the first system than that of the second system.

$$\sigma_{i,1} = 10^M \sigma_{i,2} \quad \text{Equation S2.1.5}$$

$$\ln\left(\frac{\sigma_{i,1}}{\sigma_{i,2}}\right) = \ln\left(\frac{10^M \sigma_{i,2}}{\sigma_{i,2}}\right) = \ln(10^M) \approx \frac{H_{mig,2} - H_{mig,1}}{kT} \quad \text{Equation S2.1.6}$$

$$\ln(10^M) = \ln(10) * \log_{10} 10^M \approx 2.30M \approx \frac{H_{mig,2} - H_{mig,1}}{kT} \qquad \text{Equation S2.1.7}$$

According to Figure 5 of Brett et al.,[4] a two order of magnitude change in ionic conductivity corresponds to allowing an electrolyte layer to be 100 times as thick. Therefore, a decrease in oxygen migration barrier of about 0.4 eV is equivalent to allowing an electrolyte layer to be 100 times as thick.

### S2.2. Effect of change in migration barrier on cathodic ASR.

In this section, we derive the effect of a change in migration barrier on the area specific resistance (ASR) of a perovskite material using Adler-Lane-Steele theory[5] and an empirical relationship between oxygen tracer diffusion coefficient $D^*$ and surface exchange coefficient $k$.[6] Using the assumptions of similarity given in the introduction of Section S2, we find that lowering the migration barrier by 0.6 eV reduces the ASR by a factor of 100

We set Equation S2.2.1, Equation S2.2.2, and Equation S2.2.3 to be the same as Equations 18, 23, and 24, respectively, from Adler et al.,[5] where $R_{chem}$ is the ASR, $R$ is the universal gas constant, $T$ is temperature, $F$ is Faraday's constant, $\tau$ is the solid-phase tortuosity, $\epsilon$ is porosity, $k$ (also sometimes given as $k^*$) is the surface exchange coefficient, $D^*$ is the tracer diffusion coefficient, $D_v$ is the vacancy diffusion coefficient, $c_v$ is the vacancy concentration, $c_{mc}$ is the concentration of oxygen sites, $r_0$ is the exchange-neutral flux density, $a$ is the surface area, and $\alpha$ are exchange reaction constants.

$$R_{chem} = \left(\frac{RT}{2F^2}\right)\sqrt{\frac{\tau}{(1-\epsilon)c_V D_V a r_0 (\alpha_f + \alpha_b)}} \qquad \text{Equation S2.2.1}$$

$$r_0(\alpha_f + \alpha_b) = k c_{mc} \qquad \text{Equation S2.2.2}$$

$$D^* = f D_V \frac{c_V}{c_{mc}} \qquad \text{Equation S2.2.3}$$

Rearranging Equation S2.2.3 gives Equation S2.2.4. Substituting Equation S2.2.2 and Equation S2.2.4 into Equation S2.2.1 gives Equation S2.2.5. Applying the assumptions of similarity given in the introduction of Section S2 gives Equation S2.2.6.

$$c_V D_V = \frac{c_{mc}}{f} D^* \qquad \text{Equation S2.2.4}$$

$$R_{chem} = \left(\frac{RT}{2F^2}\right)\sqrt{\frac{\tau}{(1-\epsilon)\frac{c_{mc}}{f}D^* a k c_{mc}}} \qquad \text{Equation S2.2.5}$$

$$\frac{R_{chem,1}}{R_{chem,2}} \approx \sqrt{\frac{D_2^* k_2}{D_1^* k_1}} \qquad \text{Equation S2.2.6}$$

We write the ASR of the first system as some orders of magnitude $M$ greater than the ASR of the second system and take the base-10 logarithm of both sides to give Equation S2.2.7. For example, if $R_{chem,1}$ is two orders of magnitude greater than $R_{chem,2}$, that is, $R_{chem,1}$ = 100*$R_{chem,2}$, then $M$=2. Equation S2.2.7 simplifies to Equation S2.2.8.

$$\log_{10}\frac{R_{chem,1}}{R_{chem,2}} = \log_{10}\frac{10^M * R_{chem,2}}{R_{chem,2}} = \log_{10} M = M$$

$$\Rightarrow M \approx \log_{10}\left(\frac{D_2^* k_2}{D_1^* k_1}\right)^{1/2} = \frac{1}{2}\log_{10}\left(\frac{D_2^* k_2}{D_1^* k_1}\right) \qquad \text{Equation S2.2.7}$$

$$2M \approx \log_{10}\left(\frac{D_2^* k_2}{D_1^* k_1}\right) \qquad \text{Equation S2.2.8}$$

Taking an estimate for the empirical relationship between D* and k from deSouza[6] gives Equation S2.2.9. Integrating this equation gives Equation S2.2.10. Separating the logarithm expression and dropping the base-10 subscript in Equation S2.2.8 gives Equation S2.2.11.

Substituting Equation S2.2.10 into Equation S2.2.11 gives Equation S2.2.12.

$$\frac{\partial \log_{10} k}{\partial \log_{10} D^*} \approx 0.5 \qquad \text{Equation S2.2.9}$$

$$\log_{10} k \approx 0.5 \log_{10} D^* + C \qquad \text{Equation S2.2.10}$$

$$2M \approx \log D_2^* + \log k_2 - (\log D_1^* + \log k_1) \qquad \text{Equation S2.2.11}$$

$$2M \approx \log D_2^* + 0.5 \log D_2^* + C_2 - (\log D_1^* + 0.5\log D_1^* + C_1) \qquad \text{Equation S2.2.12}$$

Assuming that the relationship intercepts $C$ are equivalent for similar systems gives Equation S2.2.13. Simplifying Equation S2.2.13 gives Equation S2.2.14. Converting to natural logarithm gives Equation S2.2.15, and dividing on both sides gives Equation S2.2.16.

$$2M \approx 1.5 \log D_2^* - 1.5 \log D_1^* \qquad \text{Equation S2.2.13}$$

$$\frac{4M}{3} \approx \log_{10}\left(\frac{D_2^*}{D_1^*}\right) \qquad \text{Equation S2.2.14}$$

$$\frac{4M}{3} \approx \log_{10}(e) \ln\left(\frac{D_2^*}{D_1^*}\right) \approx 0.4343 \ln\left(\frac{D_2^*}{D_1^*}\right) \qquad \text{Equation S2.2.15}$$

$$3.07M \approx \ln\left(\frac{D_2^*}{D_1^*}\right) \qquad \text{Equation S2.2.16}$$

Using Equation S2.2.3 and the assumptions above for correlation factor, vacancy concentration, and oxygen site concentration gives Equation S2.2.17. Using an Arrhenius expression for $D_v$, the assumptions above to cancel out the prefactors $A$, and the realization that the activation energy $Q$ for vacancy migration is the migration barrier $H_{mig}$,[7] gives Equation S2.2.18. Substituting Equation S2.2.18 into Equation S2.2.17 and then into Equation S2.2.16 gives Equation S2.2.19 and subsequently Equation S2.2.20, which relates changes in ASR to changes in migration enthalpy. From Equation S2.2.20, it is straightforward to show that if the ASR of the first system is 2 orders of magnitude larger than the ASR of the second system, or $M=2$, then the migration barrier of the first system is about 0.56 eV higher than that of the second system at 1073K (800°C), as described in the main text.

$$\frac{D_2^*}{D_1^*} = \frac{f_2 \frac{c_{V,2}}{c_{mc,2}} D_{V,2}}{f_1 \frac{c_{V,1}}{c_{mc,1}} D_{V,1}} \approx \frac{D_{V,2}}{D_{V,1}} \qquad \text{Equation S2.2.17}$$

$$\frac{D_{V,2}}{D_{V,1}} = \frac{A_2 exp(-Q_2/kT)}{A_1 exp(-Q_1/kT)} = \frac{A_2 exp(-H_{mig,2}/kT)}{A_1 exp(-H_{mig,1}/kT)} \approx \frac{exp(-H_{mig,2}/kT)}{exp(-H_{mig,1}/kT)} \qquad \text{Equation S2.2.18}$$

$$3.07M \approx \ln\left(\frac{exp(-H_{mig,2}/kT)}{exp(-H_{mig,1}/kT)}\right) \qquad \text{Equation S2.2.19}$$
$$= \ln\left(exp(-H_{mig,2}/kT)\right) - \ln\left(exp(-H_{mig,1}/kT)\right)$$

$$3.07M \approx (H_{mig,1} - H_{mig,2})/kT \qquad \text{Equation S2.2.20}$$

## S3. Additional descriptors for oxygen migration barrier

We present plots of additional descriptors for oxygen migration barrier as figures in the Figures section. For information on each descriptor, please refer to the figure caption.

The descriptors are divided into the following categories and are named Figure S3.<category letter>#:

### S3.A. Simple B-site cation properties

These descriptors are simple properties of B-site cations, including atomic number, modified Mendeleev number (see below), Shannon crystal radius, radii from A-O and B-O nearest-neighbor distances, and magnetic moment.

B-site cation polarizability was also considered,[8-12] but an examination of Table III of Shannon[13] or Table III of Shannon and Fischer[14] shows no correlation, see $Ga^{3+}$ and $Sc^{3+}$ compared for example with $Cr^{3+}$ and $V^{3+}$. Also, polarizabilities are not available for all cations, even using methods of adding polarizability. [11, 12]

In the modified Mendeleev number scheme, Mendeleev numbers are taken from Figure 1 of Ref. [15], and then Mendeleev numbers 17-33 (Ytterbium through Lanthanum, including Scandium and Yttrium) are inserted between Mendeleev number 81 (Gallium) and 82 (Lead). Ytterbium through Lanthanum become modified Mendeleev numbers 82-98, while Lead and the following elements become modified Mendeleev numbers 99 through 120. This modification allows non-radioactive elements with predominantly 3+ oxidation states to all be grouped together (e.g. $Ga^{3+}$ and $Sc^{3+}$ are now near each other instead of separated by the transition metal block), in order to reflect the similar oxygen migration behavior calculated in these perovskites. The d-block transition metal elements are also more spread out.

Shannon crystal radii[16] were taken as explained in the main text. A-site Shannon crystal radii are not shown as a separate descriptor, since there is only a single A-site value regardless of B-site cation.

Additionally, A-site and B-site radii were calculated using bulk A-O and B-O average nearest-neighbor (NN) bond distances, respectively, and assuming a fixed Shannon crystal radius of 1.26 Å for oxygen.[16] B-O distances were averaged using 6 NNs, corresponding to octahedral coordination. A-O distances were averaged using 8 NNs; although the formal A-site coordination in a perovskite is 12, shifts in A-site placements and oxygen positions resulted in large errors when averaged over 12 NNs. These errors were greatly reduced when averaged over only the closest 8 NNs. Rather than being shown as separate descriptors, these NN-based radii are shown compared to the Shannon crystal radii; these comparisons reveal that the NN-based radii are similar to the Shannon crystal radii, and would not make a significant difference in the correlations seen for radius-based descriptors. Radii from Bader charge analysis[17] were also explored but were more complex to calculate and did not yield significant differences in either radius correlations or radius-derived structural property correlations, and are not shown.

### S3.B. Structural properties from the bulk

These descriptors are structural properties that could be derived from the bulk, including average B-O-B bond angle, formula unit volume, free volume, percent free volume, Goldschmidt tolerance factor,[8] and Kilner critical radius.[18] A hard-sphere approximation is used for volume descriptors.

The B-O-B average bond angle was taken as a measure of structural ideality.[19]

Shannon crystal radii[16] were used, as discussed in the main text and in Section S3.A.

### S3.C. Defected structural properties

These descriptors are defected structural properties, including estimated fractional travel distance, fractional travel distance in the NEB calculation, and fractional vacancy crowding.

Fractional travel distance in the NEB, which is derived from the activated state, is included only for comparison to estimated fractional travel distance, which is derived only from the initial state.

These two defected structural properties were the most successful descriptors out of hundreds of bulk and initial state structural property candidates, including various distances, fractional distances, and angles evaluated for $A^{3+}B^{3+}O_3$ perovskites.

### S3.D. Bulk electronic and bonding properties

These descriptors are bulk electronic and bonding properties, including bulk cohesive energy,[20, 21] bulk modulus, Fermi energy, B-site $e_g$ electron occupation, bandgap,[22] and energy above convex hull for a closed system.

Note that electron occupation (Figure S3.D4 and Figure S3.D5), while showing a good trend, is much more complicated to extract and parse correctly than oxygen p-band center energy. Correct occupation numbers may depend heavily on the radius of the projected sphere (here, 1 Angstrom is used as the radius in all cases) and on the rotation used, where internal octahedral distortions like differing B-O bond lengths and imperfect B-site centering[23] may render an exact rotation difficult to achieve.

The cohesive energy was taken as the energy from VASP minus the spin-polarized energy of the collection of atoms using PAW pseudopotentials; that is, a spin-polarized calculation for each species in a large box of vacuum, with the resulting energy multiplied by the number of atoms of that species being used.[24]

The bulk modulus was taken using a series of (P, V) points and the third-order Birch Murnaghan equation[25] as in Ref.[3]. Although this descriptor requires additional calculations, we include it because the calculations are short and because bulk modulus could presumably also be determined or tabulated from experiment.

### S3.E. Descriptors from the main paper

These descriptors are descriptors from the main paper, reprinted in larger format, which are Mendeleev number, vacancy formation energy, and oxygen p-band center energy.

## S4. Range of migration and vacancy formation energies

The range for oxygen migration barriers in charge-compensated and doped systems is estimated at 0.4 eV based on studies of all hops in three systems. The ranges for charge-compensated systems in Ref.[3] ESI Figure S8.3 and Figure S8.4 were approximately 0.25 eV. Here, Figure S4.1 shows a range of 0.4 eV for charge-compensated $LaScO_3$. The range for uncompensated $LaScO_3$, in Figure S4.2, is slightly higher at 0.5 eV, but only charge-compensated systems are shown in the main text, so the 0.4 eV range is used in the main text. Because we do not know where in the range a particular hop will fall unless all hops are calculated, the range bars in Figure 1 of the main text are shown as +/- 0.4 eV for systems where there is not data for all hops.

Figure S4.3 shows that dopant positioning over three systems, using two hops and three dopant positions, had a largest spread in migration barriers of 0.3 eV. This spread is lower than the spread for hop directions discussed above. Therefore, for doped systems, we continue to use the more conservative +/- 0.4 eV range.

Oxygen vacancy formation energies had a maximum range of 0.06 eV over all 24 oxygen sites within each of three charge-compensated $LaXO_3$ sytems (see Figure S4.4). However, doped systems had a larger range, both among three oxygen sites with fixed dopant positioning and among the three oxygen sites with three different dopant positions. The largest spread in vacancy formation energies was 0.4 eV (see Figure S4.5). Therefore, the range value of +/- 0.4 eV for oxygen vacancy formation energies is used in Figure 3 in the main text.

## Tables

## Table S1.1. Literature values for Figure 1

Table S1.1. Literature values for Figure 1 in the main paper. Activation energy as calculated from an Arrhenius plot for the vacancy diffusion coefficient $D_v$ or for the ionic conductivity $\sigma_i$ of Sr-doped LaM$^{III}$O$_{3-\delta}$, which is equivalent to migration barrier enthalpy $H_{mig}$.[7, 26, 27] See ESI Section S9 of Mayeshiba and Morgan for the applicability of comparing our $E_{mig}(V)$ with $H_{mig}(P)$.[3]

| Actual material | Temp. (°C) | $H_{mig}$ (eV) | Notes | Reference |
|---|---|---|---|---|
| LaAlO$_3$, weakly acceptor doped | 600-800 | 1.00 | Activation energy for $D_v$, reported on Ref.[28] Figure 15. | 28 |
| La$_{0.9}$Sr$_{0.1}$Al$_{0.9}$Mg$_{0.1}$O$_{3-\delta}$ | 800 | 0.9 | Activation energy for $\sigma_i$ as assumed from $\sigma_{total}$ in reducing atmosphere (Lybye et al, p.98) [29] | 29 |
| La$_{0.9}$Sr$_{0.1}$Sc$_{0.9}$Mg$_{0.1}$O$_{3-\delta}$ | 800 | 0.5 | Activation energy for $\sigma_i$ as assumed from $\sigma_{total}$ in reducing atmosphere (Lybye et al, p.98) [29] | 29 |
| La$_{0.9}$Sr$_{0.1}$ScO$_{3-\delta}$ | 730-980 | 0.71 | Activation energy for $\sigma_{total}$ in N$_2$, which is expected to be $\sigma_i$ (Nomura and Tanase, p. 234) [26] | 26 |
| La$_{0.9}$Sr$_{0.1}$ScO$_{3-\delta}$ | 330-480 | 0.47 | Activation energy for $\sigma_{total}$ in N$_2$, which is expected to be $\sigma_i$ (Nomura and Tanase, p. 234) [26] | 26 |
| BaTiO$_3$, weakly acceptor doped | 200-900 | 0.70 | Activation energy for $D_v$, reported on Ref.[28] Figure 15. | 28 |
| SrTiO$_3$, weakly acceptor doped | 150-1400 | 0.65 | Activation energy for $D_v$, reported on Ref.[28] Figure 15. | 28 |
| La$_{0.7}$Ca$_{0.3}$CrO$_3$ | 900-1000 | 0.81 | Activation energy for $D_v$ | 30 |
| La$_{0.79}$Sr$_{0.20}$MnO$_{3-\delta}$ | 700-860 | 0.726 | Activation energy from chemical diffusion coefficient $\widetilde{D}$ [a] | 31 |
| La$_{0.8}$Sr$_{0.2}$MnO$_3$ | 850-1000 | 1.47 | Activation energy for $D_v$ converted from $\widetilde{D}$ (De Souza and Kilner, Fig. 8, line B, and Yasuda and Hishinuma, Fig. 10) [32, 33] | 32, 33 |
| $LaFeO_{3-\delta}$ | 900-1100 | 0.767 | Activation energy for $D_v$ | 27 |
| $La_{0.9}Sr_{0.1}FeO_{3-\delta}$ | 850-1100 | 0.819 | Activation energy for $D_v$ | 27 |
| $La_{0.75}Sr_{0.25}FeO_{3-\delta}$ | 900-1050 | 1.182 | Activation energy for $D_v$; authors note that large activation energy may come from inaccuracy in $D_0^*$ (Ishigaki et al. 1988, p. 184) [27] | 27 |
| $LaCoO_{3-\delta}$ | 800-1000 | 0.798 | Activation energy for $D_v$ | 27 |
| $La_{0.9}Sr_{0.1}CoO_{3-\delta}$ | 800-1000 | 0.819 | Activation energy for $D_v$ | 27 |
| $LaCoO_3$ | 850-1000 | 0.781 | Activation energy for $D_v$ | 7 |
| $La_{0.9}Sr_{0.1}Ga_{0.9}Mg_{0.1}O_{3-\delta}$ | 200, 800 | 1.2 | Activation energy for $\sigma_i$ as assumed from $\sigma_{total}$ in reducing atmosphere (Lybye et al., p.98). [29] Temperature is given as 200°C on Lybye et al., p.98 and 800°C in Lybye et al., Table 5. [29] | 29 |
| $La_{0.9}Sr_{0.1}Ga_{0.9}Mg_{0.1}O_{3-\delta}$ | 1000 | 0.6 | Activation energy for $\sigma_{total}$, expected to be almost purely ionic (Lybye et al., p.99, p.101) [29] | 29 |
| $La_{0.9}Sr_{0.1}GaO_{3-\delta}$ | 730-980 | 0.6 | Activation energy for $\sigma_{total}$ in $N_2$, which is expected to be $\sigma_i$ (Nomura and Tanase, p. 234) [26] | 26 |
| $La_{0.9}Sr_{0.1}GaO_{3-\delta}$ | 430-580 | 0.81 | Activation energy for $\sigma_{total}$ in $N_2$, which is expected to be $\sigma_i$ (Nomura and Tanase, p. 234) [26] | 26 |
| $BaZrO_3$–based solid solution | 300-800 | 1 | Approximated over four references in Ref.[28] Figure 13 and estimated from conductivity studies; $BaZr_{0.9}Y_{0.1}O_{3-\delta}$;[34, 35] $BaZr_{0.8}Y_{0.2}O_{3-\delta}$;[36] $BaZr_{0.9}Yb_{0.1}O_{3-\delta}$.[37] | 28 |
| $La_{0.9}Sr_{0.1}In_{0.9}Mg_{0.1}O_{3-\delta}$ | 800 | 0.8-0.9 | Estimated "despite the disintegration in reducing | 29 |
| | | | atmosphere" (Lybye et al, p. 99) | |
| $La_{0.9}Sr_{0.1}InO_{3-\delta}$ | 730-980 | 0.84 | Activation energy for $\sigma_{total}$ in $N_2$, which is expected to be $\sigma_i$ (Nomura and Tanase, p. 234) [26] | 26 |
| $La_{0.9}Sr_{0.1}InO_{3-\delta}$ | 330-480 | 0.70 | Activation energy for $\sigma_{total}$ in $N_2$, which is expected to be $\sigma_i$ (Nomura and Tanase, p. 234) [26] | 26 |

[a]According to Equation 16 in Ishigaki et al., the chemical diffusion coefficient for vacancy mediated diffusion can be equated to the vacancy diffusion coefficient using $\widetilde{D} = -\frac{1}{2}\frac{1}{\left(\frac{\partial \ln C_v}{\partial \ln P(O_2)}\right)}D_v$.[27] For a similar system, the denominator of the second fraction approaches a constant (Yasuda and Hishinuma, Fig. 7)[32] at higher oxygen partial pressures. Therefore, for this case, $\widetilde{D} \propto D_v$, and on an Arrhenius plot, the activation energy of $\widetilde{D}$ would be equivalent to the activation energy of $D_v$.

## Figures

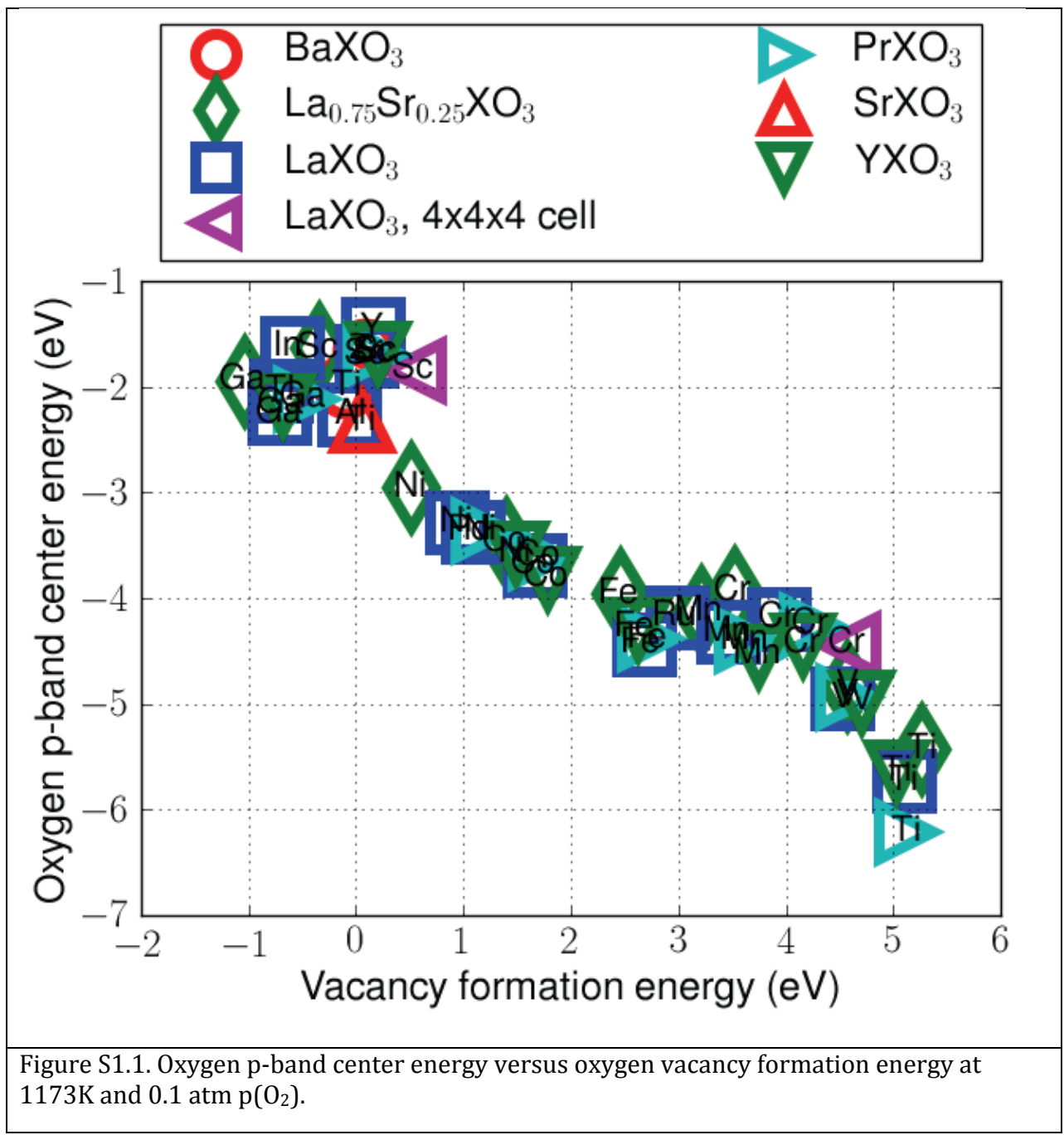

Figure S1.1. Oxygen p-band center energy versus oxygen vacancy formation energy at 1173K and 0.1 atm p(O$_2$).

Figure S1.1. Oxygen p-band center energy versus oxygen vacancy formation energy

Figure S1.2. Literature comparison of our data (plus signs) to other calculations (open symbols), which are labeled by the first letter of the author's last name: C=Ref.[38], D=Ref.[39], I=Ref.[40], K=Ref.[41], M=Ref.[42], R=Ref.[43], T=Ref.[44], and Y=Ref.[45]

Colors indicate the series to which we believe the data should be compared, if possible. The oxidation state 2+ or 3+ indicates the nominal oxidation state of a nearby B-site cation during migration for GGA and GGA+U calculations. For potentials, the choice of calculation comparison is less clear: although potentials are optimized based on 3+ cations, there is no

> comparable idea of charge compensation for an undoped system.
>
> Each of our data points should be taken within a +/- 0.4 eV range to account for symmetry-distinct hops and dopant effects (see Section S4). This range is particularly important when comparing to potential data, where additional higher-barrier or lower-barrier hops as well as dopant-vacancy association effects may have been studied.
>
> We see agreement within the range for GGA and GGA+U data. Additional discrepancy for potential data may be due to explicit charge compensation in our systems that is not present using potentials, maybe especially for single-oxidation-state 3+ cations (Chromium may count in this category as well, since it is strongly prefers the 3+ oxidation state under octahedral coordination[46]). [40]

**Figure S1.2. Literature comparison to other calculations.**

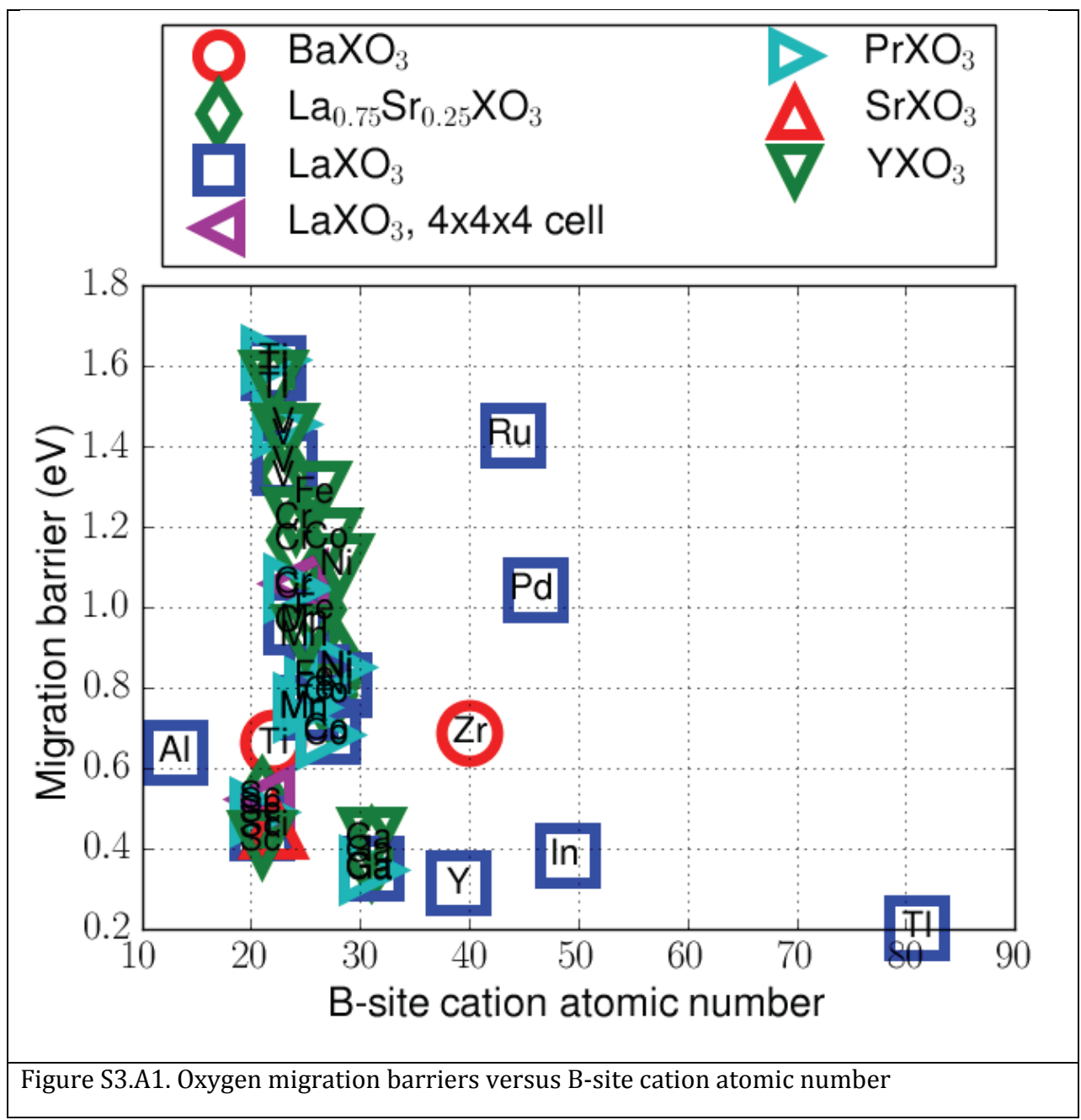

Figure S3.A1. Oxygen migration barriers versus B-site cation atomic number

**Figure S3.A1. B-site cation atomic number**

Figure S3.A2. Oxygen migration barriers versus B-site cation modified Mendeleev number (see Section S3 for an explanation).[15]

**Figure S3.A2. B-site cation modified Mendeleev number**

Figure S3.A3. Oxygen migration barriers versus B-site cation crystal radius, determined as described in Computational Methods of the main paper

Figure S3.A3. B-site cation crystal radius

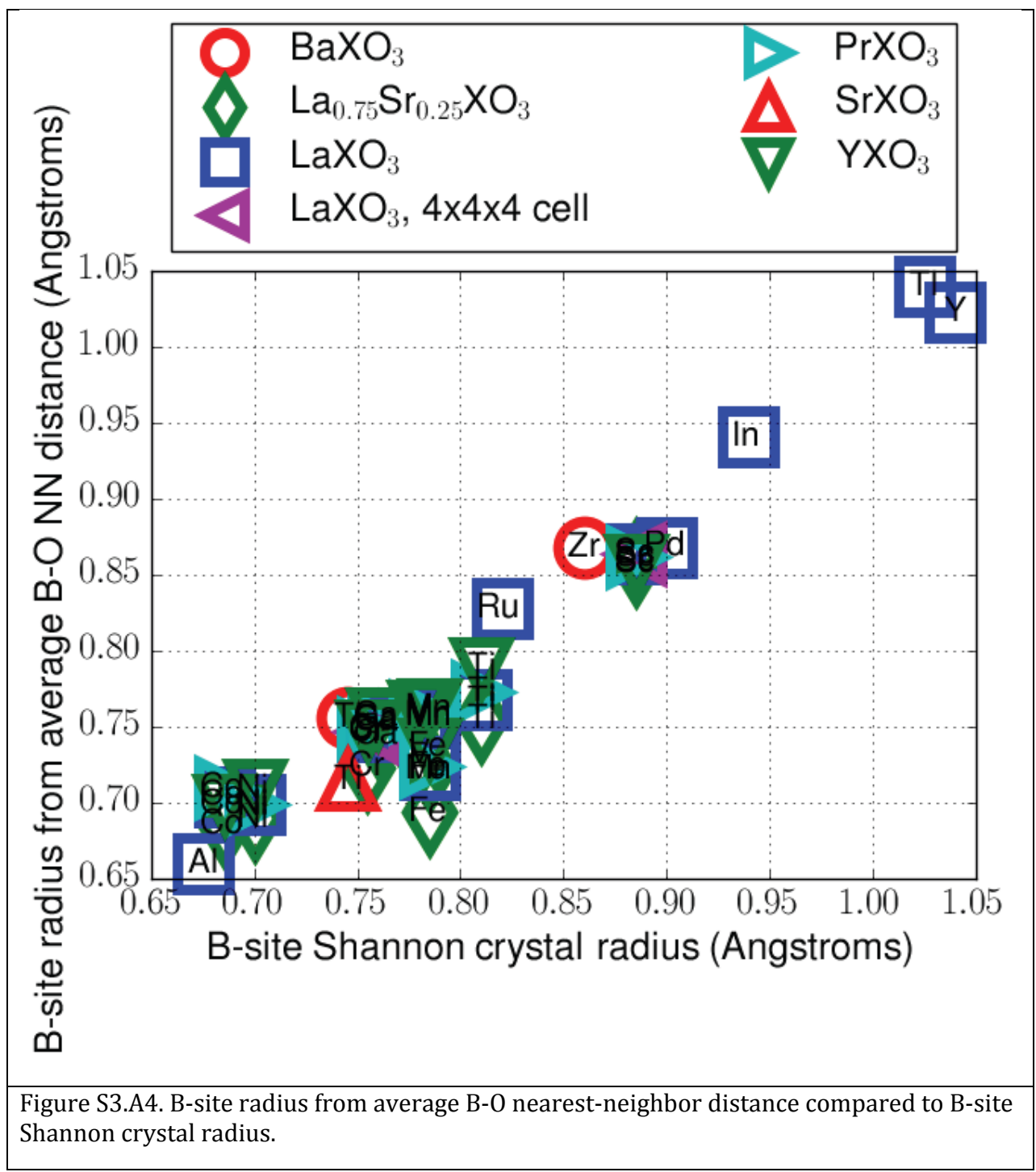

Figure S3.A4. B-site radius from average B-O nearest-neighbor distance compared to B-site Shannon crystal radius.

Figure S3.A4. B-site radius comparison

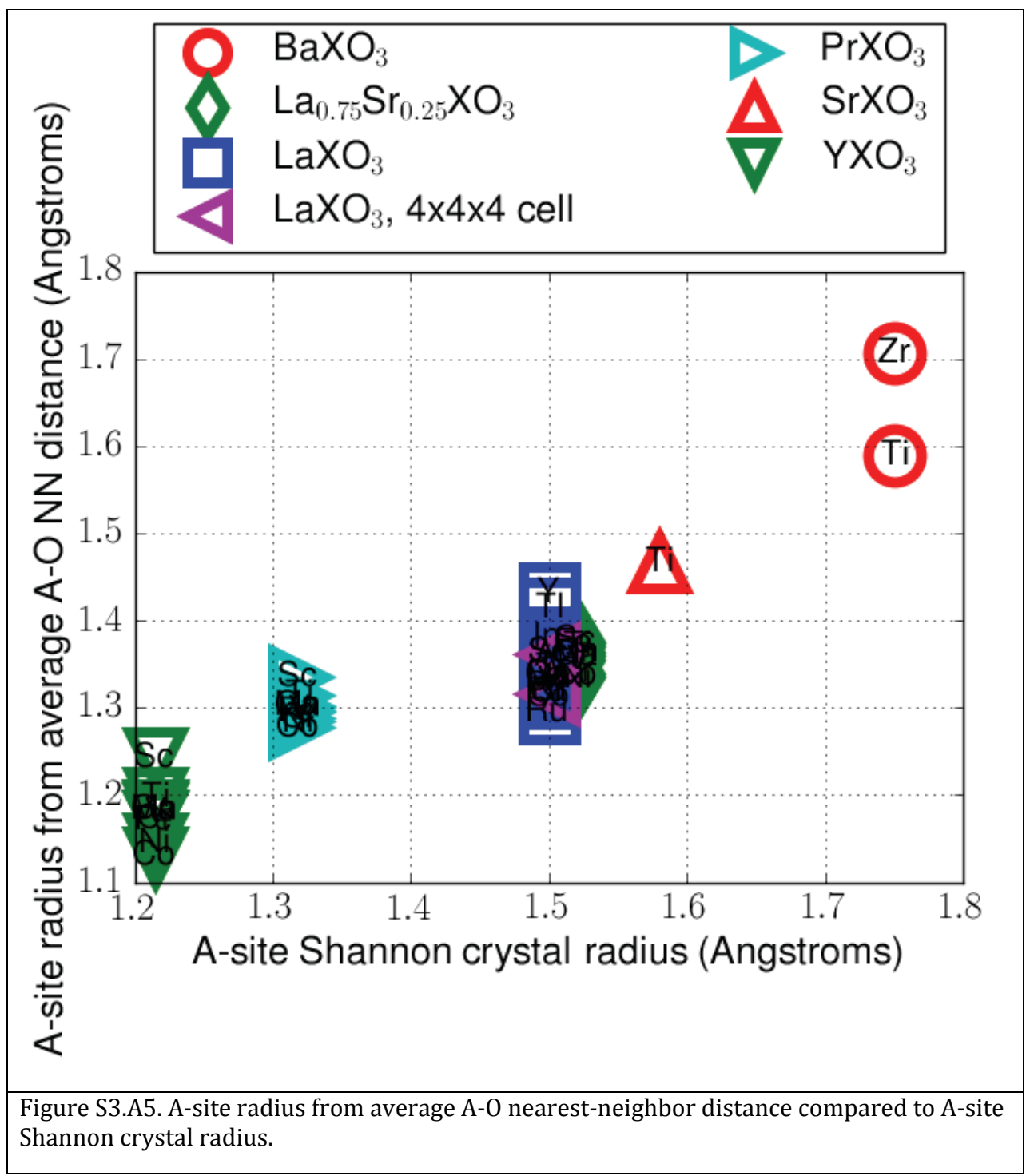

Figure S3.A5. A-site radius from average A-O nearest-neighbor distance compared to A-site Shannon crystal radius.

**Figure S3.A5. A-site radius comparison**

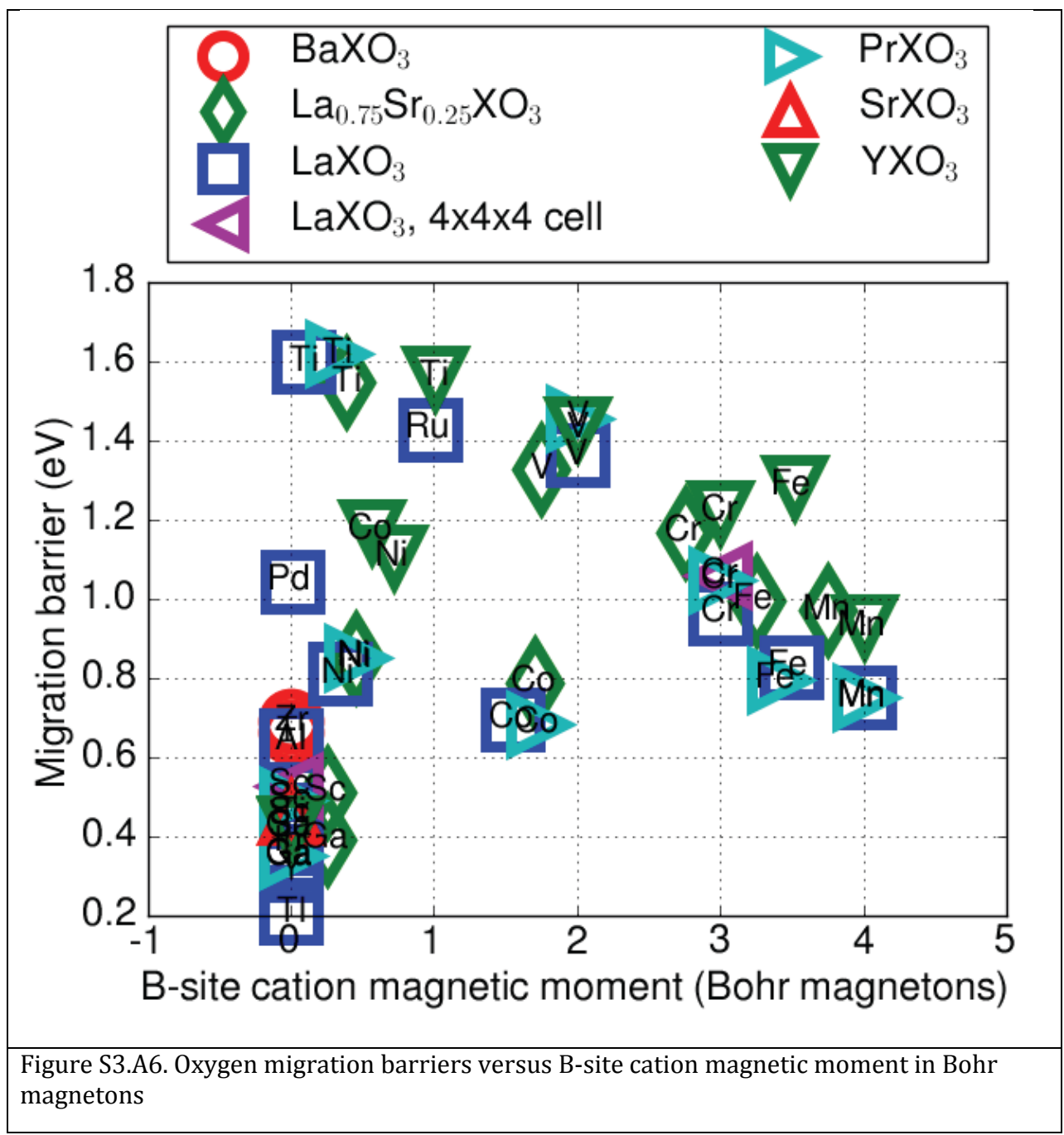

Figure S3.A6. Oxygen migration barriers versus B-site cation magnetic moment in Bohr magnetons

Figure S3.A6. B-site cation magnetic moment

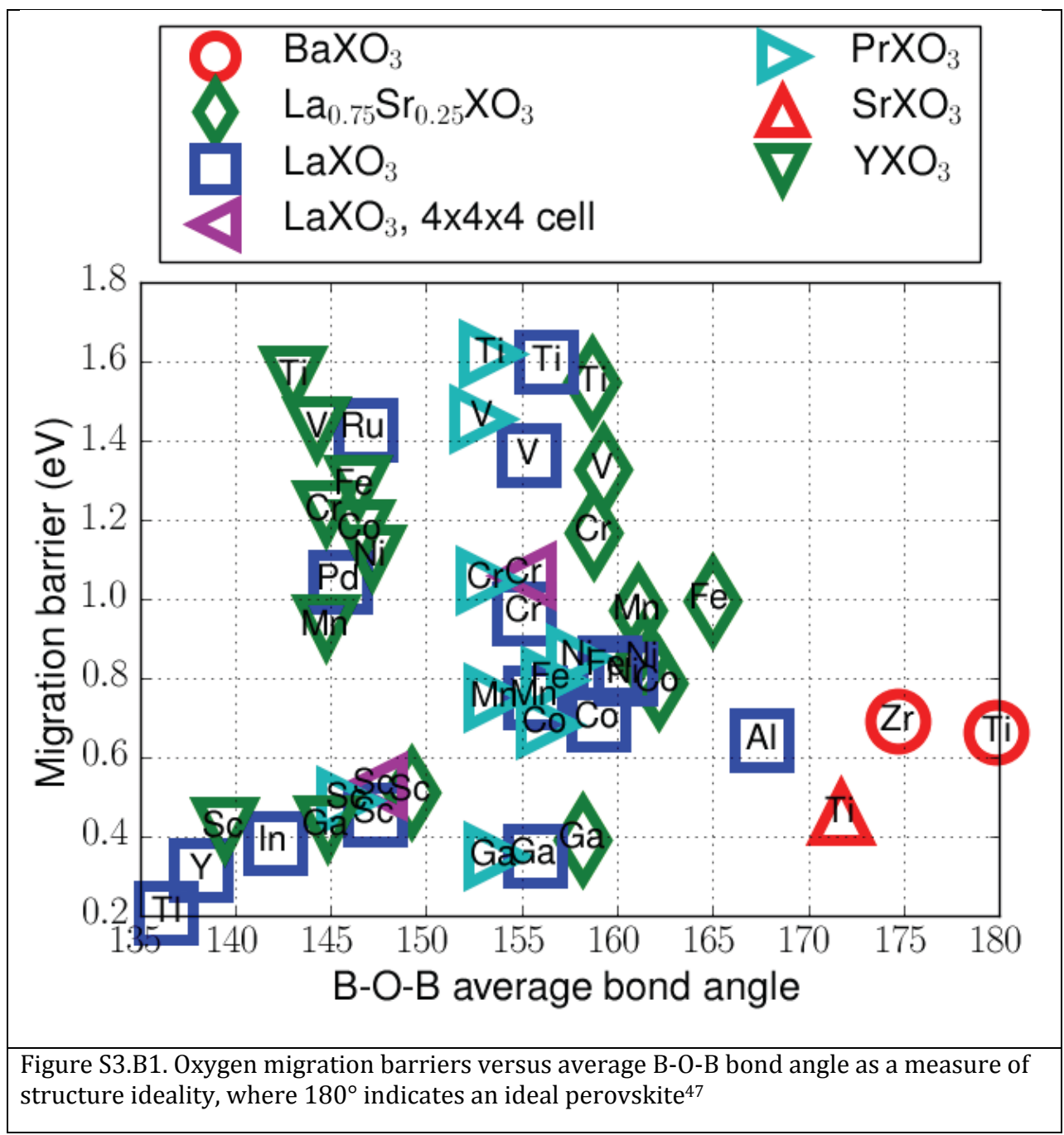

Figure S3.B1. Oxygen migration barriers versus average B-O-B bond angle as a measure of structure ideality, where 180° indicates an ideal perovskite[47]

**Figure S3.B1. Average B-O-B bond angle**

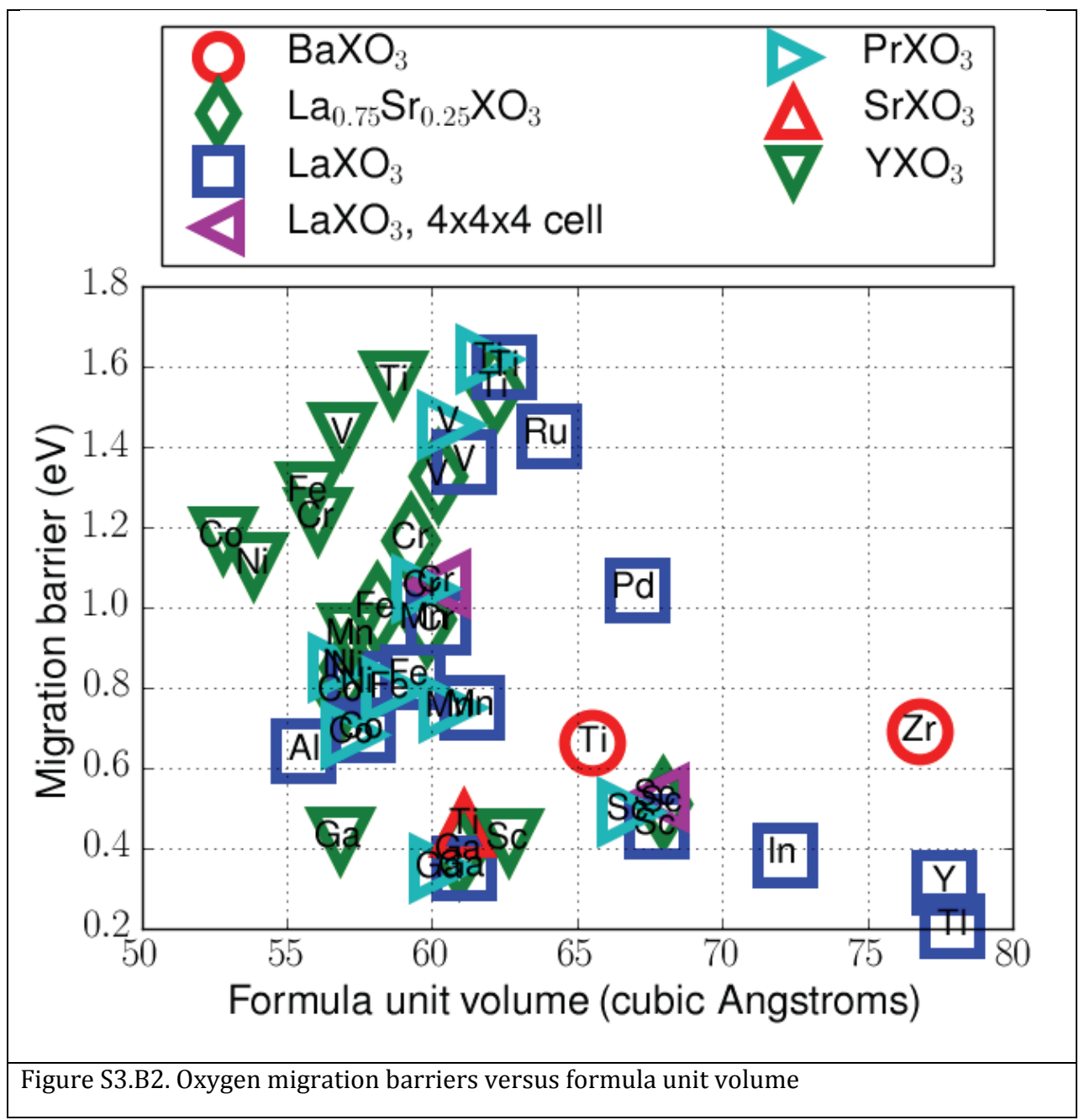

Figure S3.B2. Oxygen migration barriers versus formula unit volume

**Figure S3.B2. Formula unit volume**

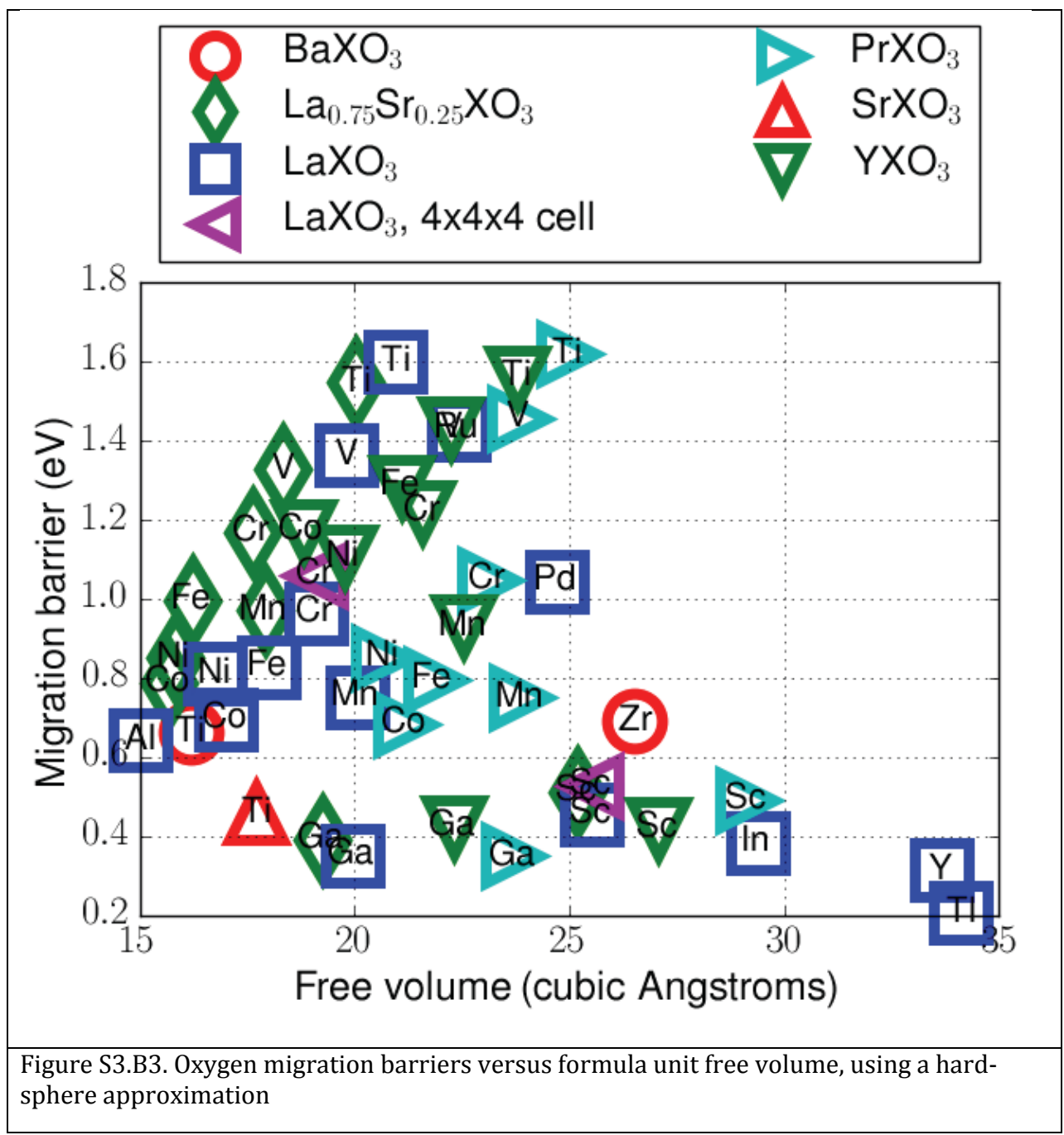

Figure S3.B3. Oxygen migration barriers versus formula unit free volume, using a hard-sphere approximation

**Figure S3.B3. Free volume**

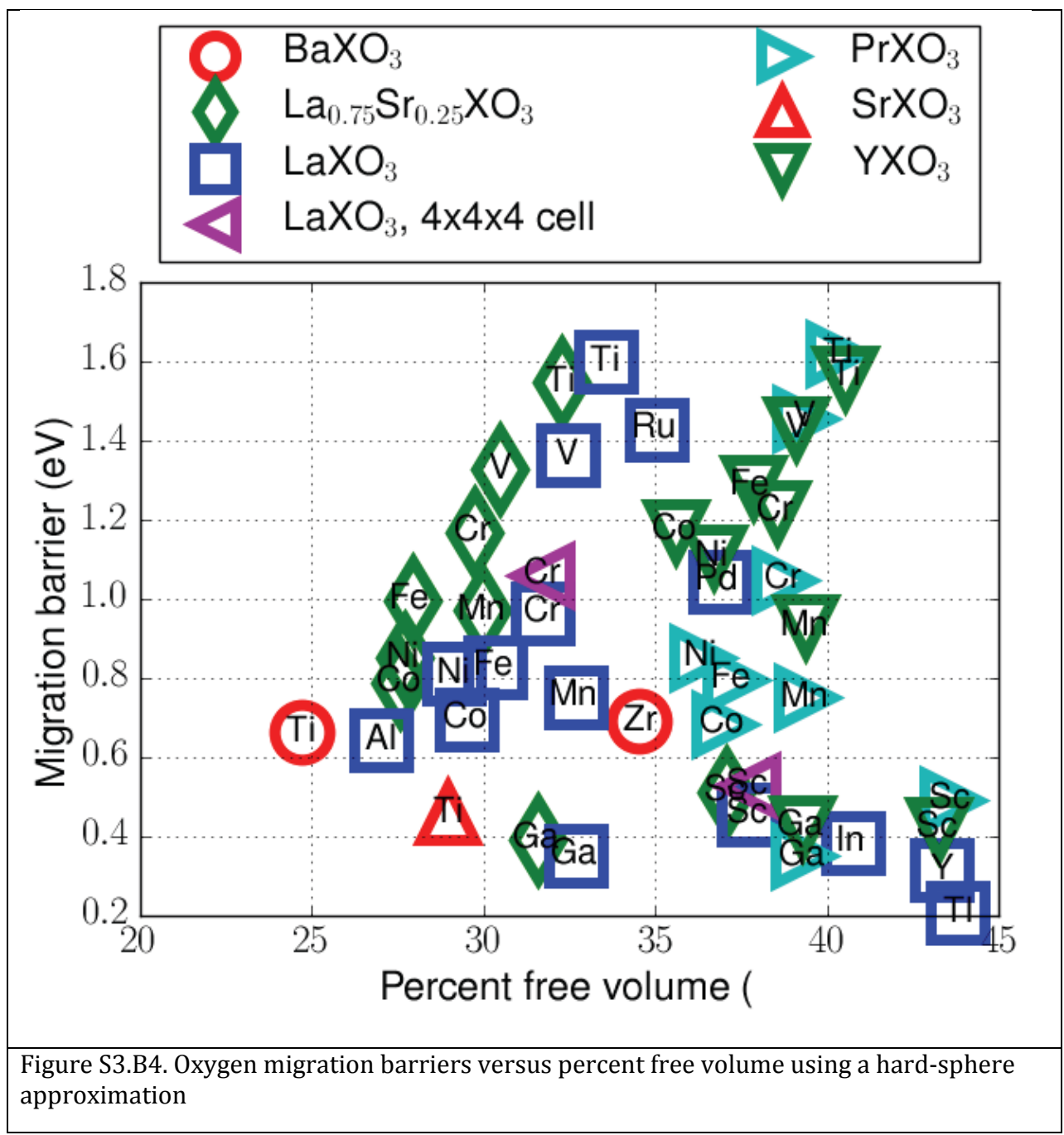

Figure S3.B4. Oxygen migration barriers versus percent free volume using a hard-sphere approximation

**Figure S3.B4. Percent free volume**

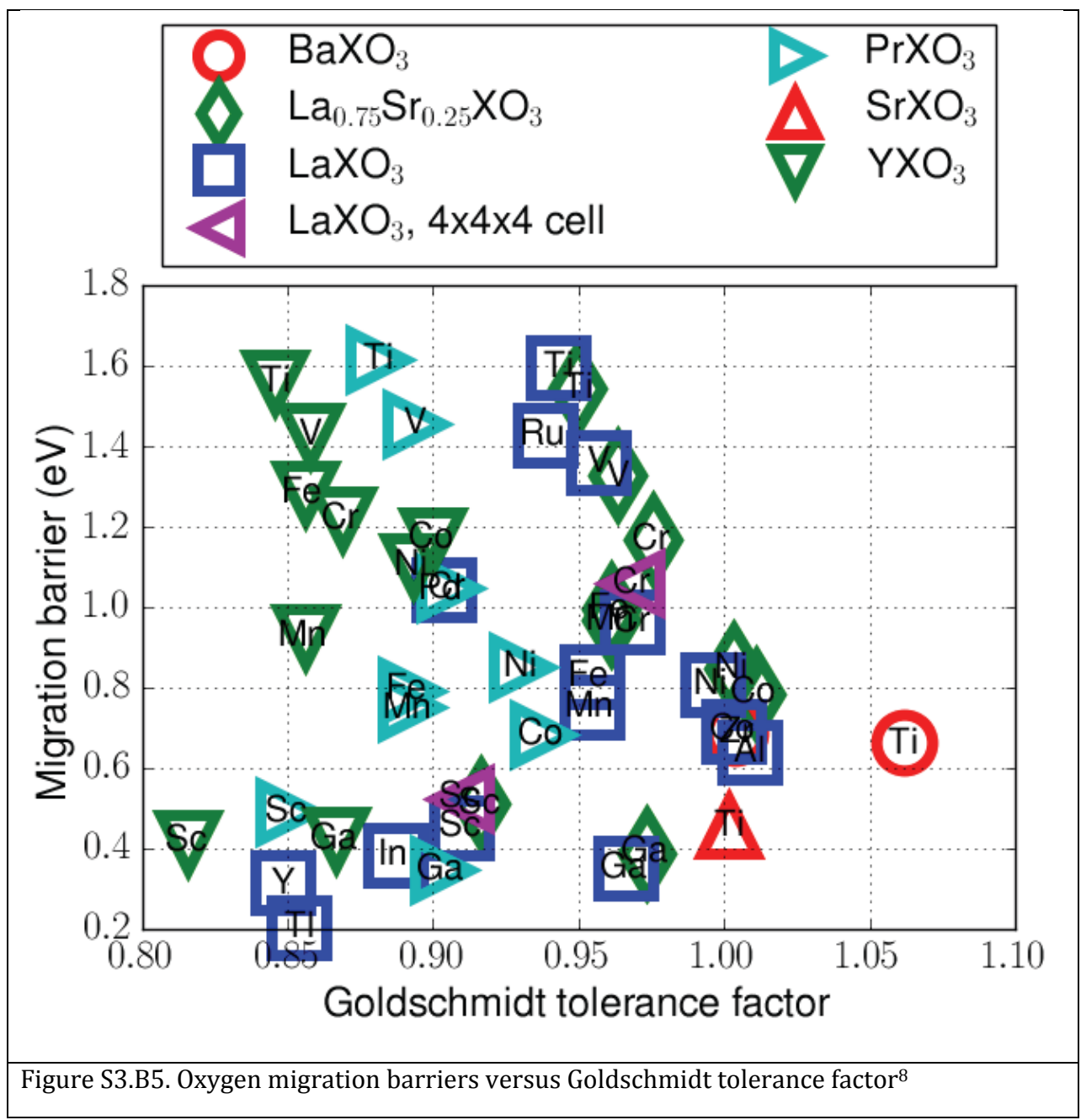

Figure S3.B5. Oxygen migration barriers versus Goldschmidt tolerance factor[8]

**Figure S3.B5. Goldschmidt tolerance factor**

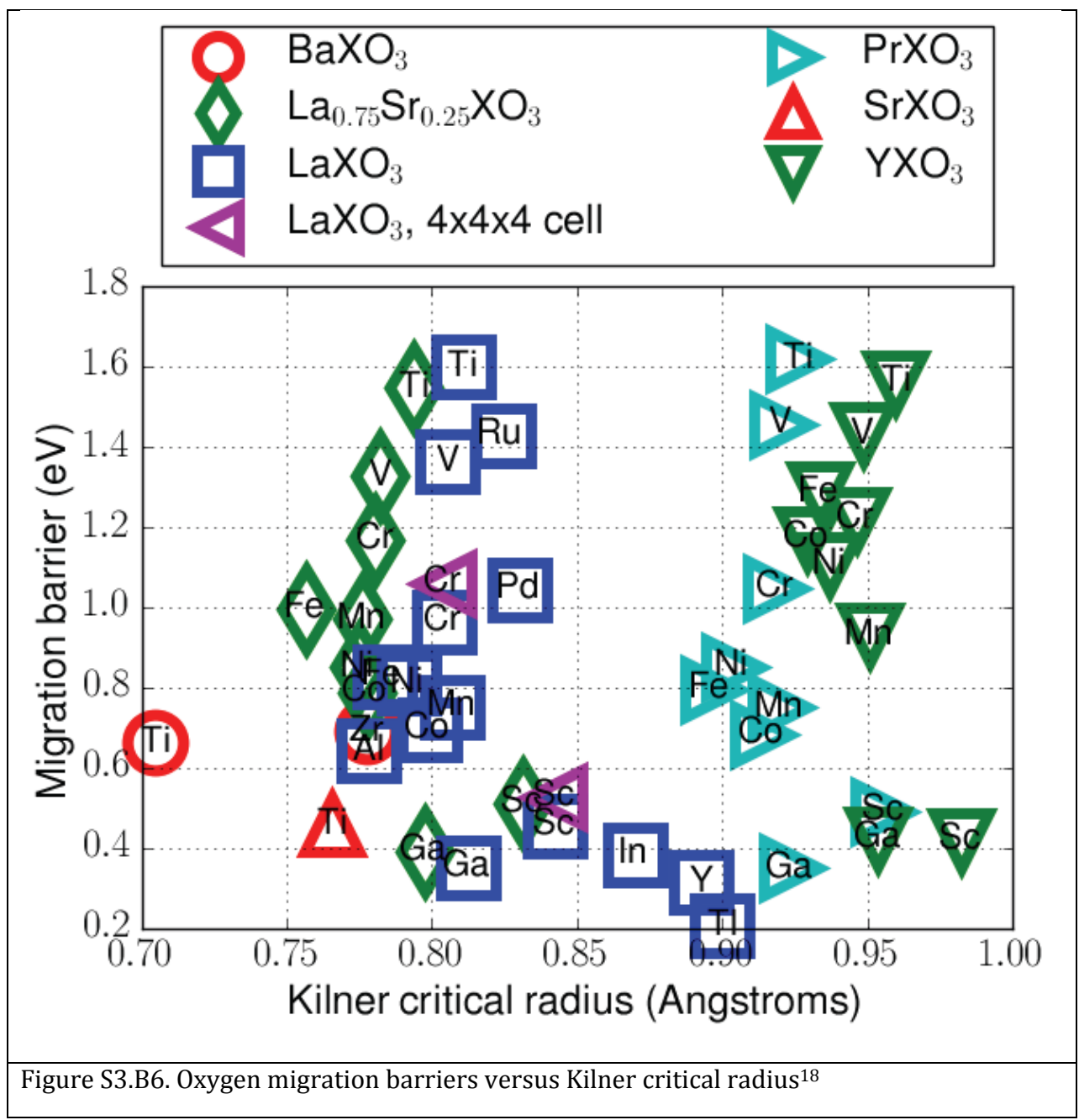

Figure S3.B6. Oxygen migration barriers versus Kilner critical radius[18]

**Figure S3.B6. Kilner critical radius**

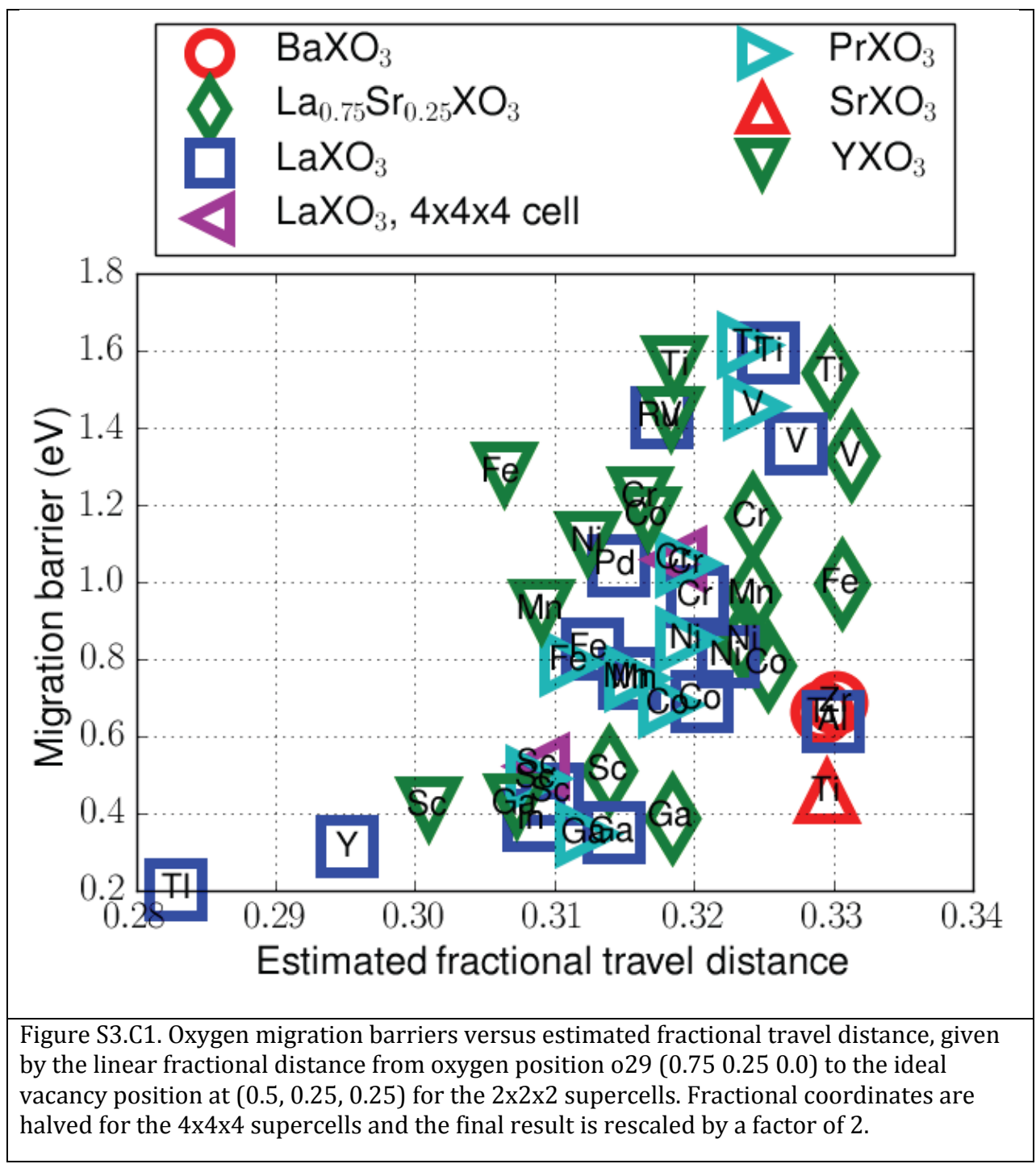

Figure S3.C1. Oxygen migration barriers versus estimated fractional travel distance, given by the linear fractional distance from oxygen position o29 (0.75 0.25 0.0) to the ideal vacancy position at (0.5, 0.25, 0.25) for the 2x2x2 supercells. Fractional coordinates are halved for the 4x4x4 supercells and the final result is rescaled by a factor of 2.

**Figure S3.C1. Estimated fractional travel distance**

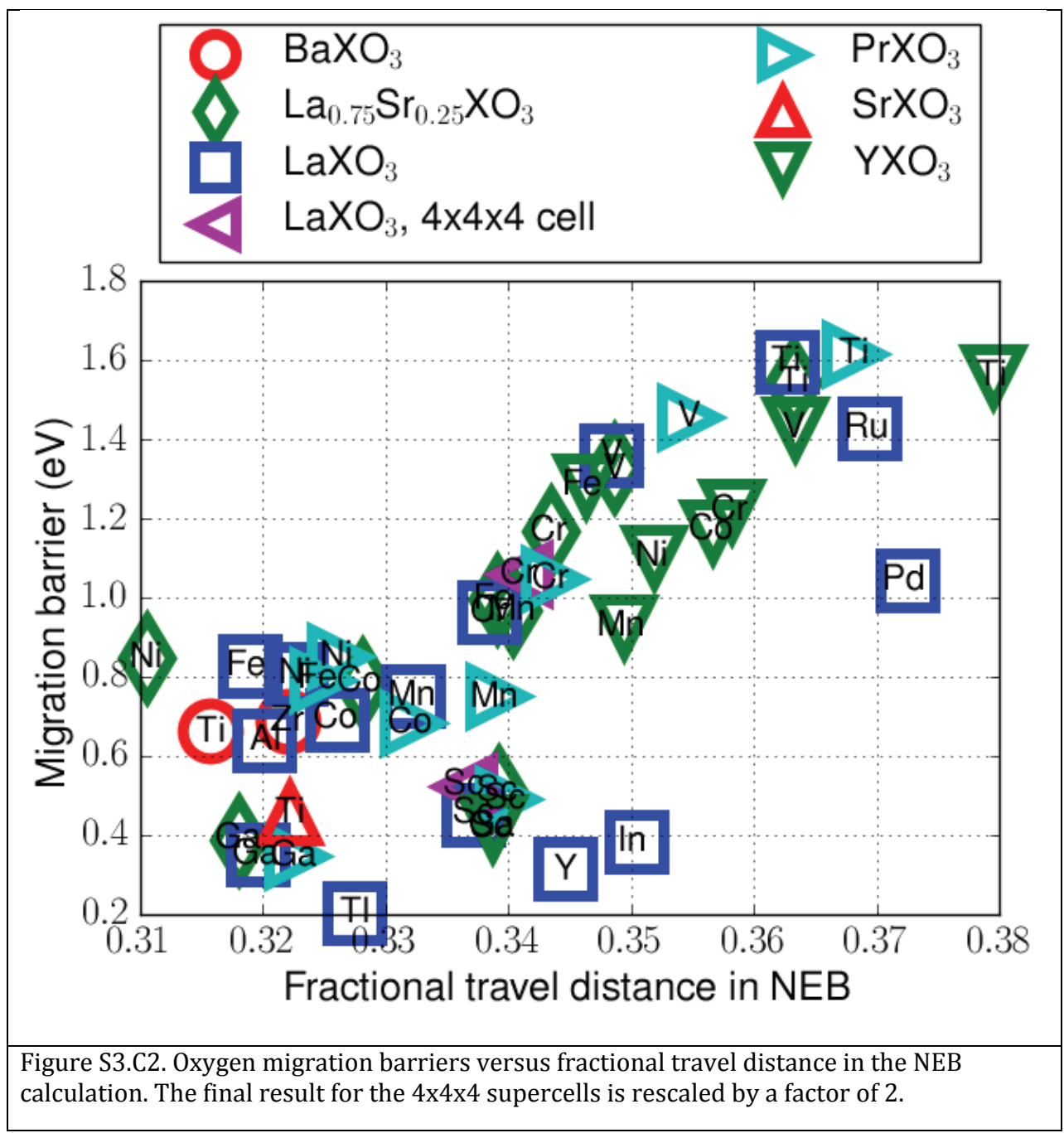

Figure S3.C2. Oxygen migration barriers versus fractional travel distance in the NEB calculation. The final result for the 4x4x4 supercells is rescaled by a factor of 2.

**Figure S3.C2. Fractional travel distance in the NEB calculation**

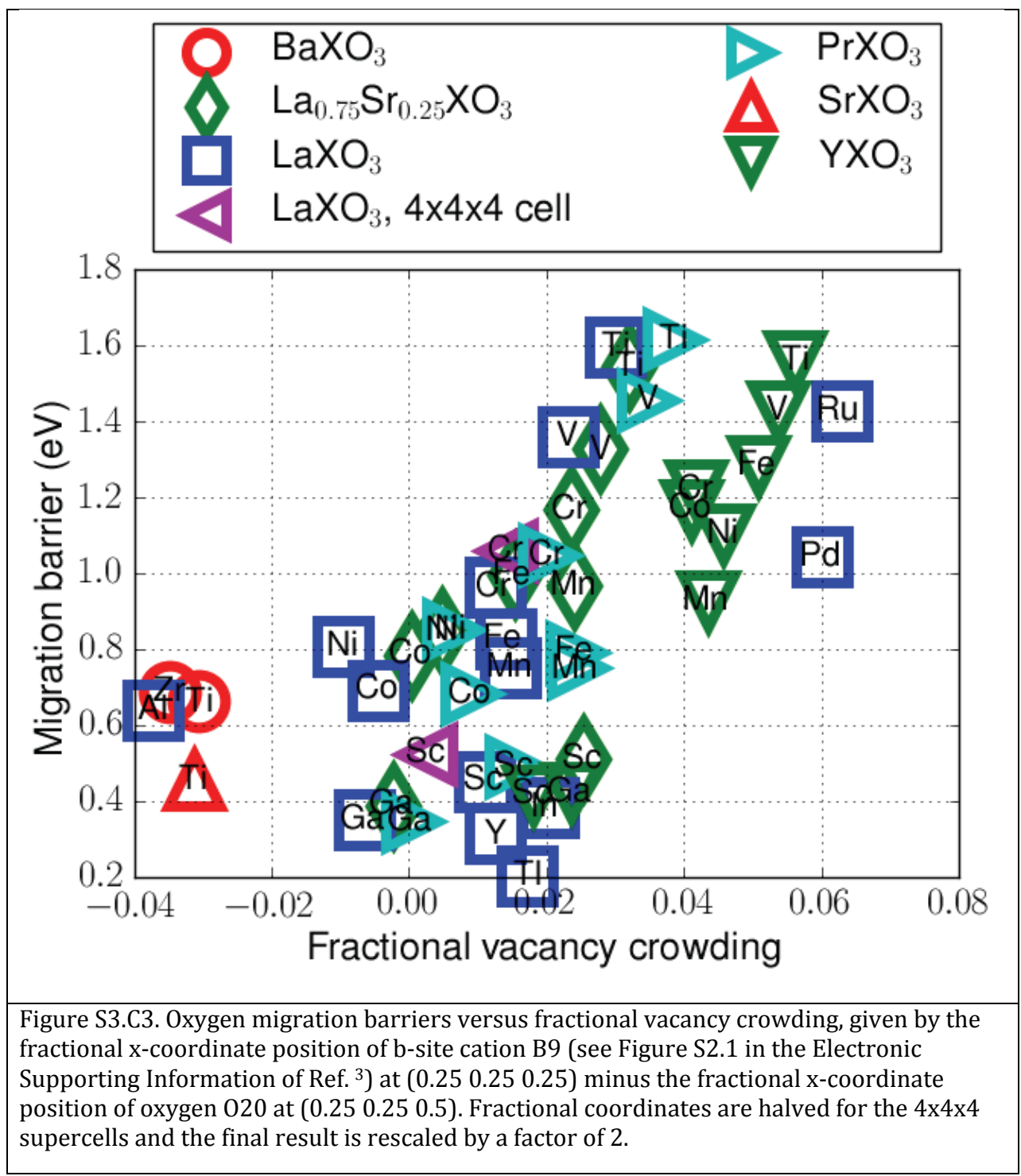

Figure S3.C3. Oxygen migration barriers versus fractional vacancy crowding, given by the fractional x-coordinate position of b-site cation B9 (see Figure S2.1 in the Electronic Supporting Information of Ref. [3]) at (0.25 0.25 0.25) minus the fractional x-coordinate position of oxygen O20 at (0.25 0.25 0.5). Fractional coordinates are halved for the 4x4x4 supercells and the final result is rescaled by a factor of 2.

**Figure S3.C3. Fractional vacancy crowding**

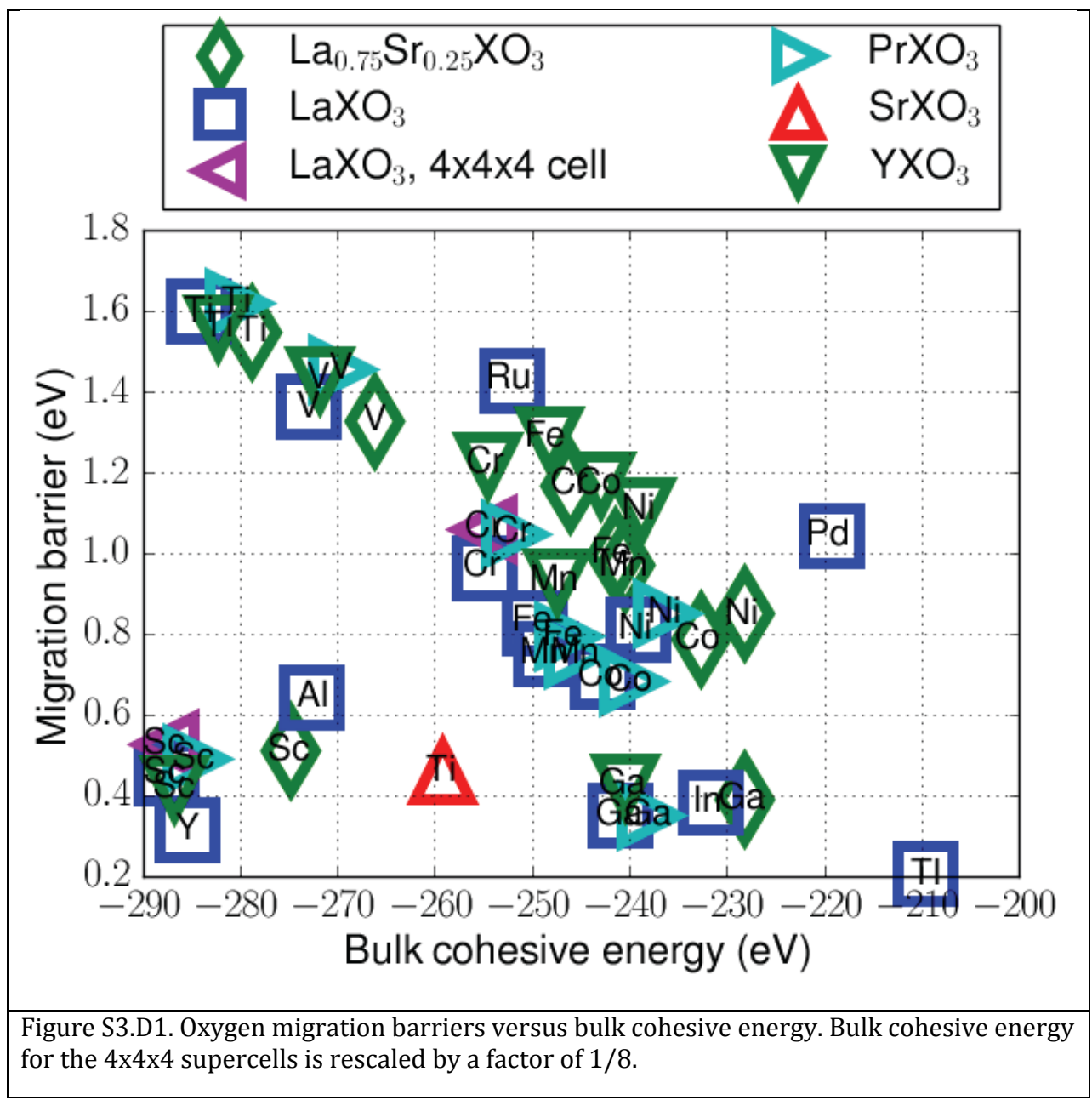

Figure S3.D1. Oxygen migration barriers versus bulk cohesive energy. Bulk cohesive energy for the 4x4x4 supercells is rescaled by a factor of 1/8.

**Figure S3.D1. Bulk cohesive energy**

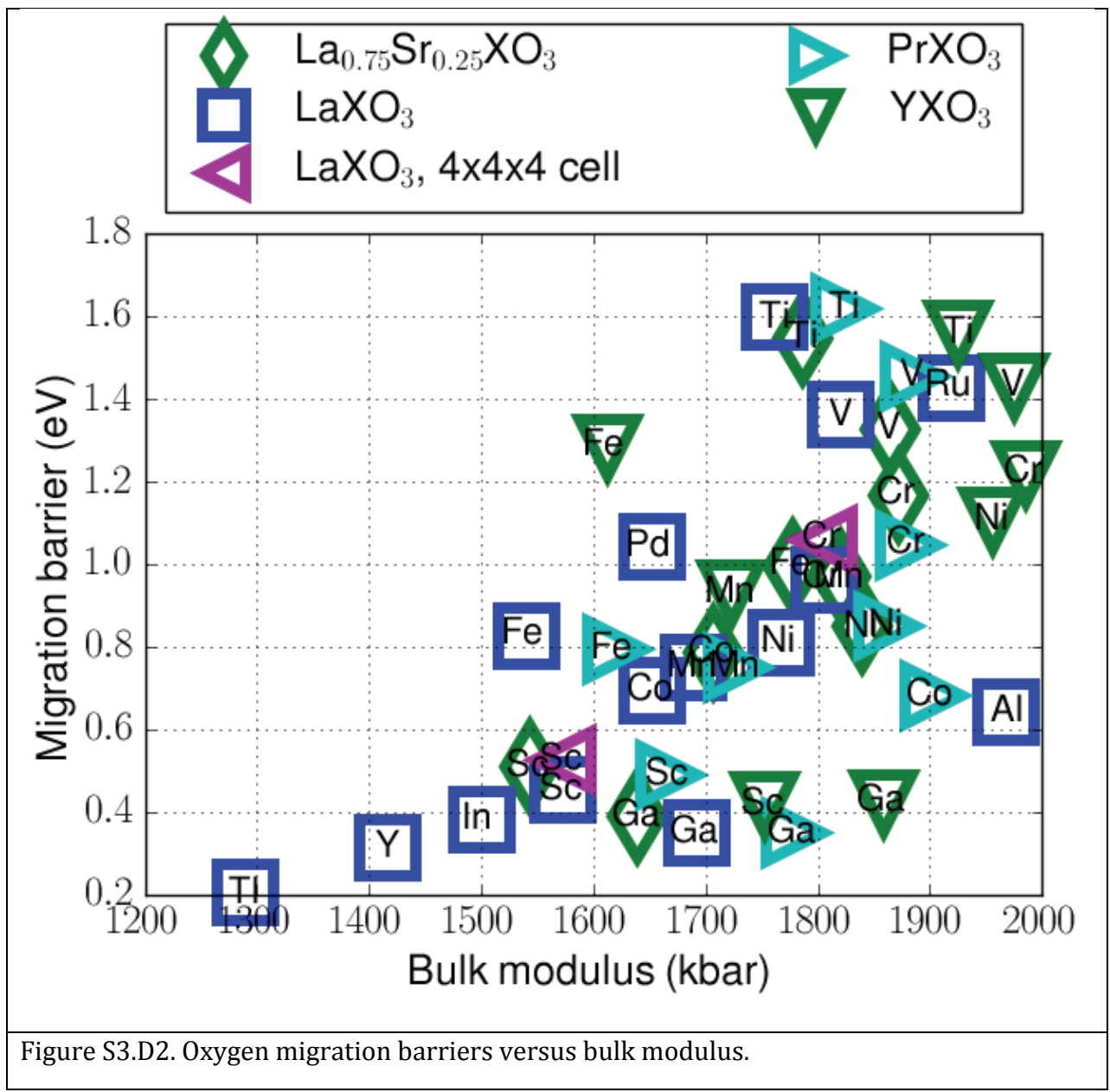

Figure S3.D2. Oxygen migration barriers versus bulk modulus.

**Figure S3.D2. Bulk modulus**

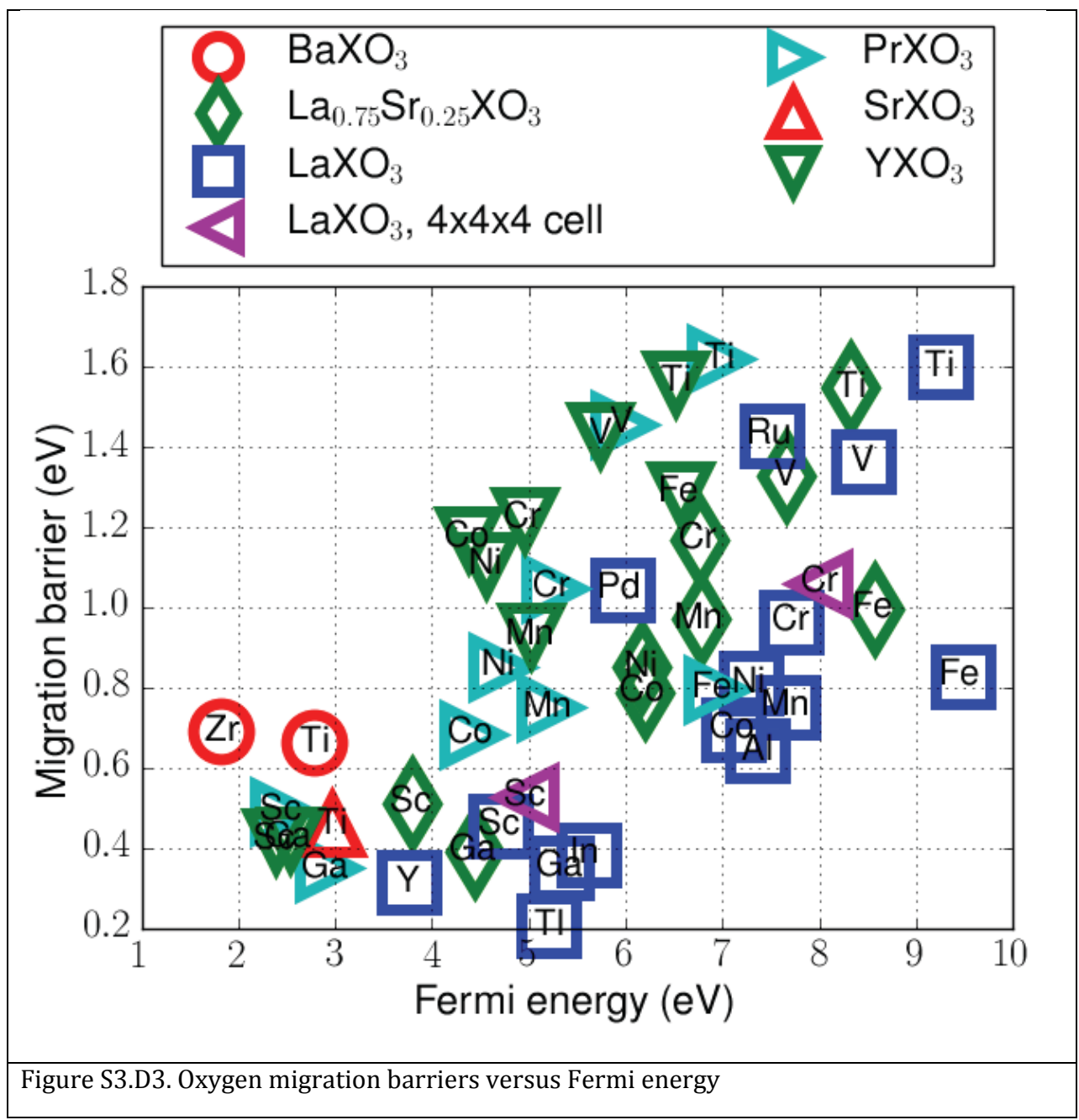

Figure S3.D3. Oxygen migration barriers versus Fermi energy

**Figure S3.D3. Fermi energy**

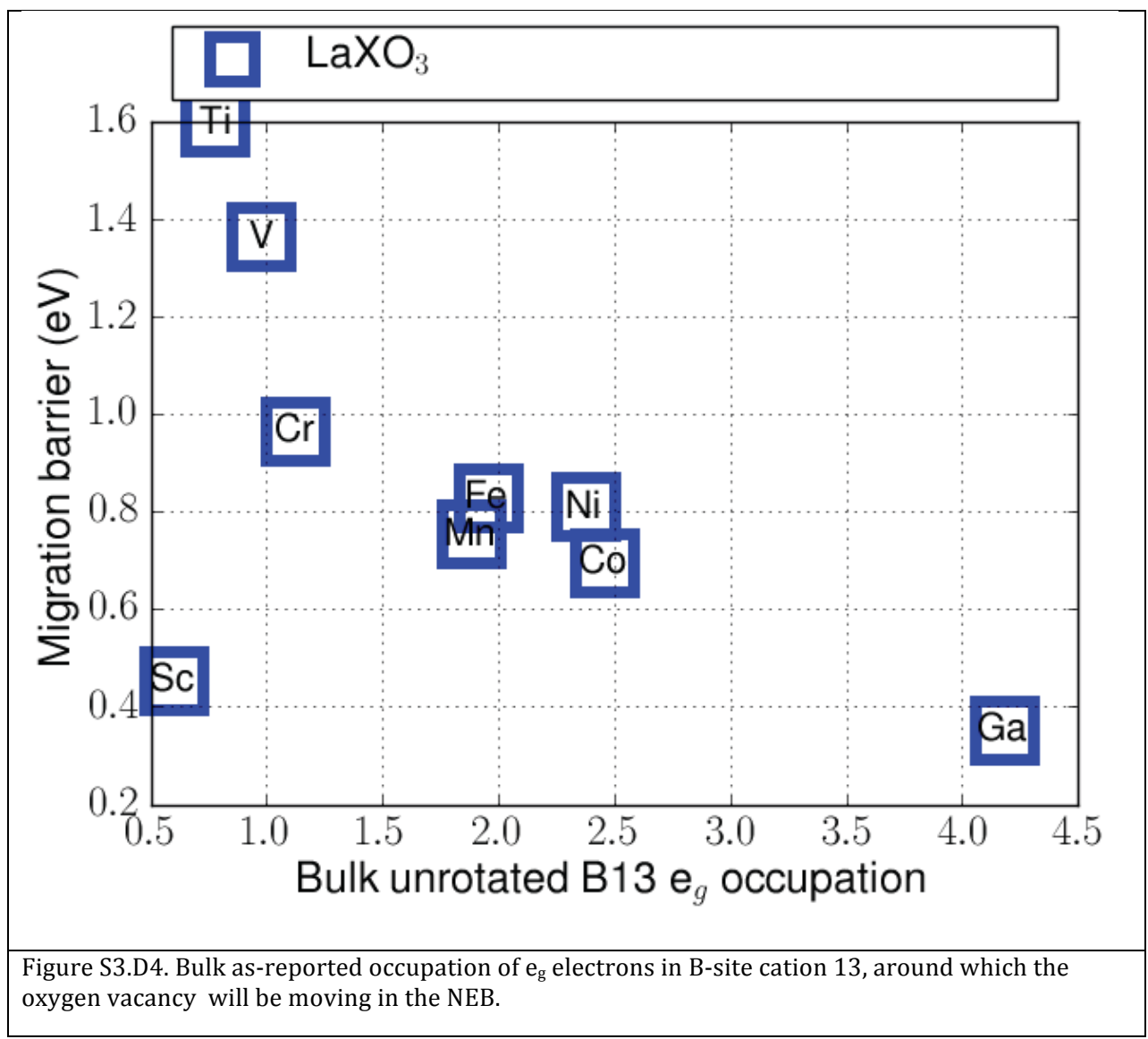

Figure S3.D4. Bulk as-reported occupation of $e_g$ electrons in B-site cation 13, around which the oxygen vacancy will be moving in the NEB.

**Figure S3.D4. Bulk as-reported occupation of $e_g$ electrons in B-site cation 13**

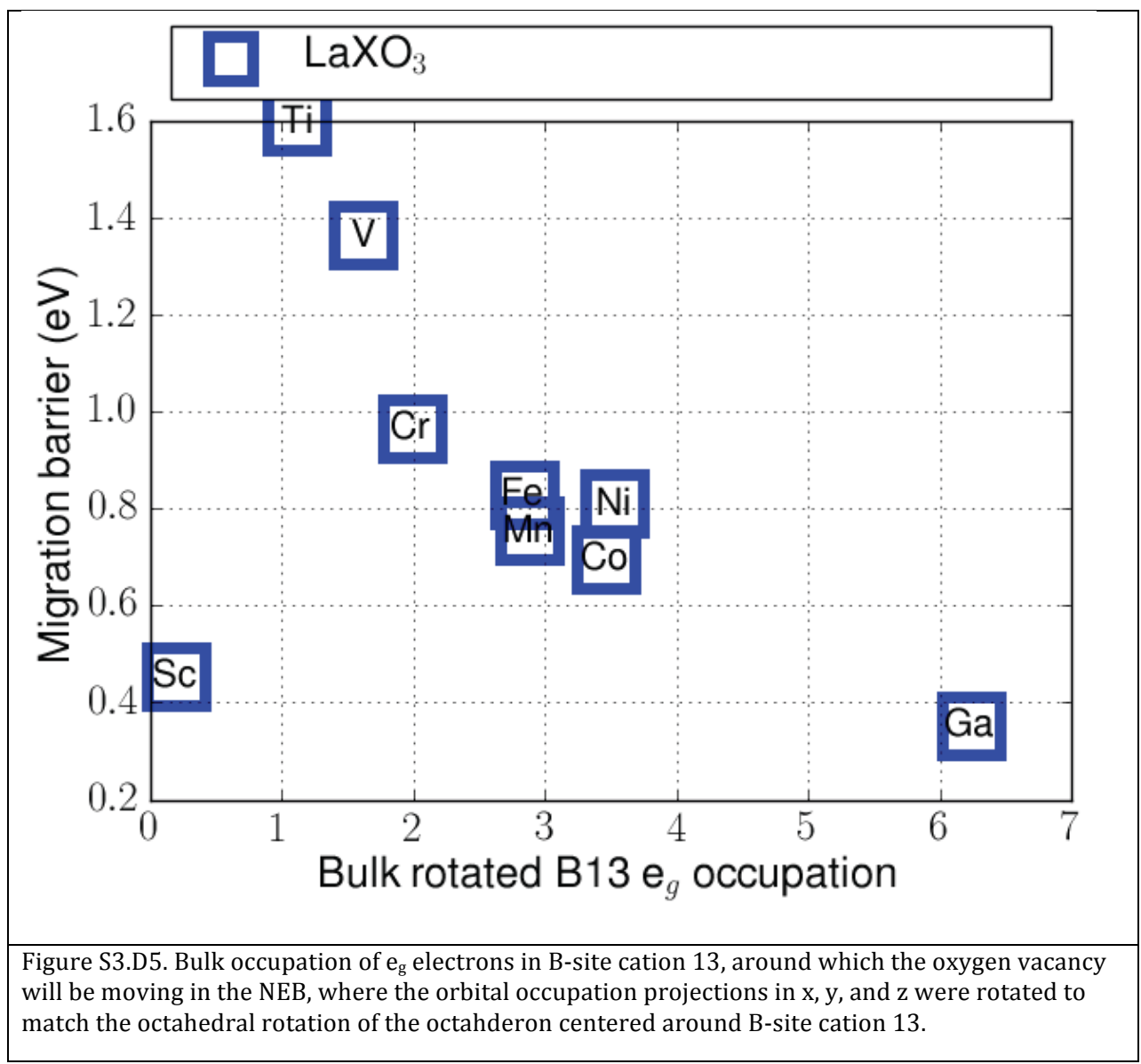

Figure S3.D5. Bulk occupation of $e_g$ electrons in B-site cation 13, around which the oxygen vacancy will be moving in the NEB, where the orbital occupation projections in x, y, and z were rotated to match the octahedral rotation of the octahderon centered around B-site cation 13.

Figure S3.D5. Bulk occupation of $e_g$ electrons in B-site cation 13, rotated

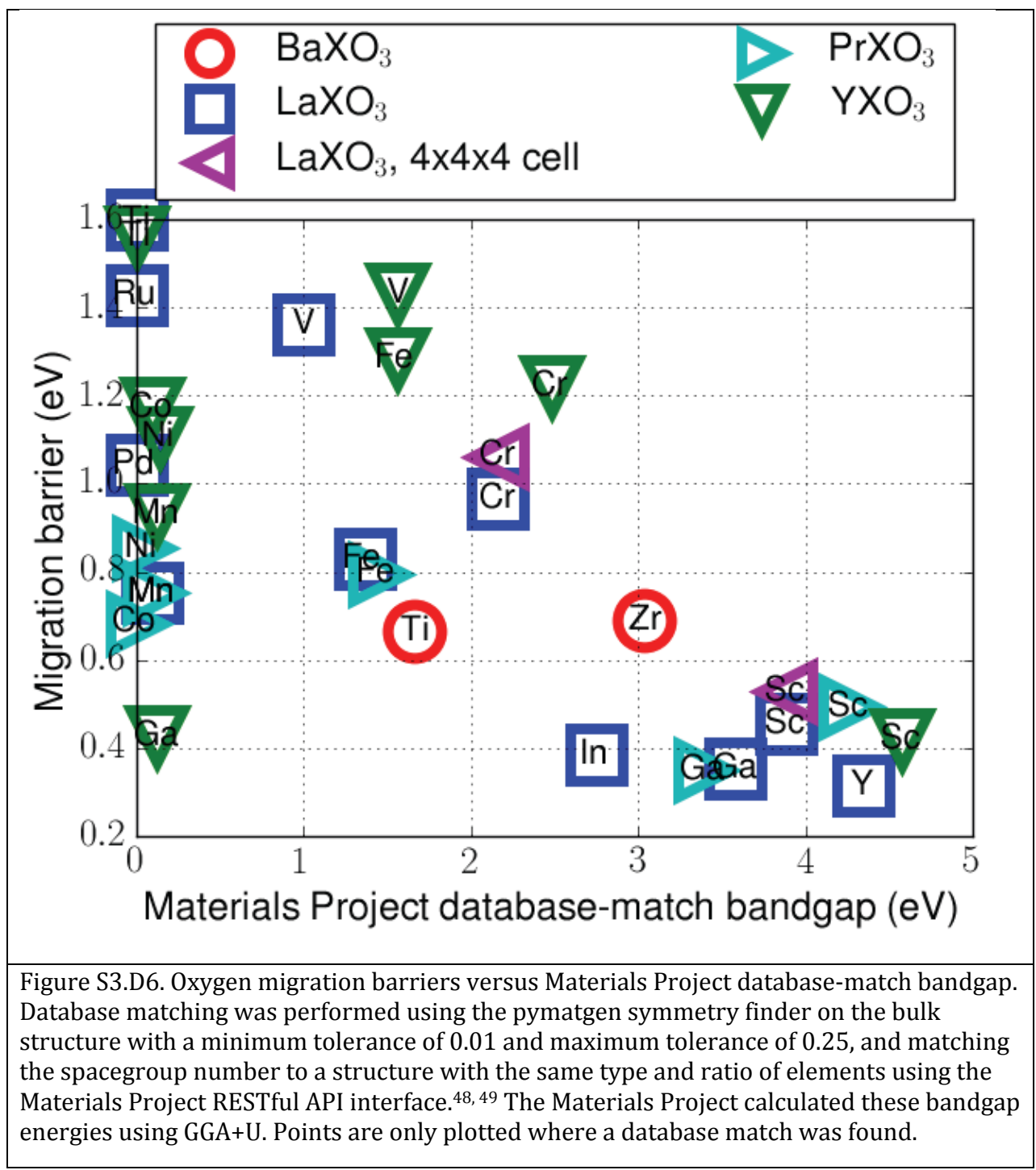

Figure S3.D6. Oxygen migration barriers versus Materials Project database-match bandgap. Database matching was performed using the pymatgen symmetry finder on the bulk structure with a minimum tolerance of 0.01 and maximum tolerance of 0.25, and matching the spacegroup number to a structure with the same type and ratio of elements using the Materials Project RESTful API interface.[48, 49] The Materials Project calculated these bandgap energies using GGA+U. Points are only plotted where a database match was found.

**Figure S3.D6. Bandgap**

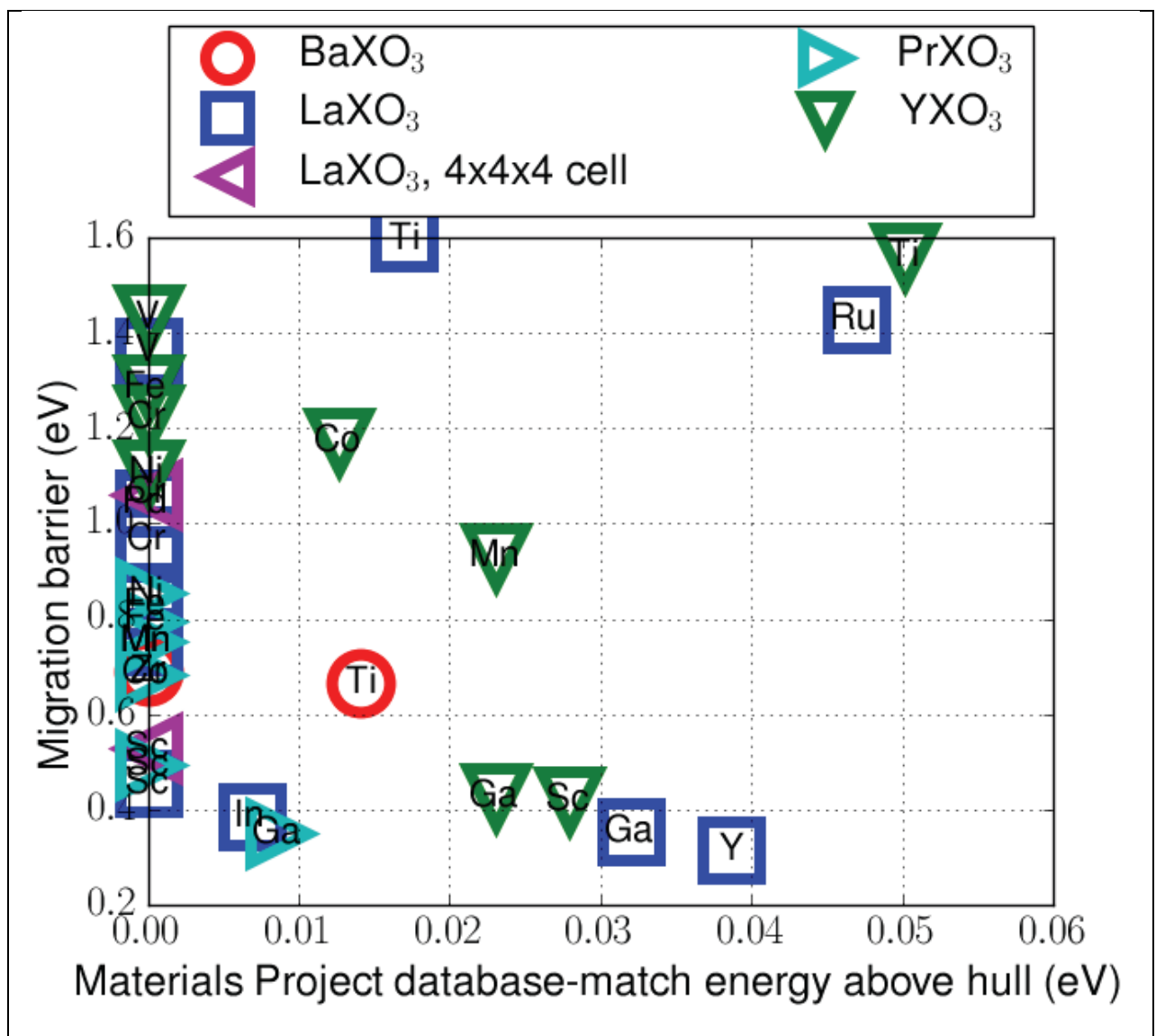

Figure S3.D7. Oxygen migration barriers versus Materials Project database-match energy above the convex hull (closed system). Database matching was performed using the pymatgen symmetry finder on the bulk structure with a minimum tolerance of 0.01 and maximum tolerance of 0.25, and matching the spacegroup number to a structure with the same type and ratio of elements using the Materials Project RESTful API interface.[48, 49] The Materials Project calculated the energies for their phase diagrams using GGA+U. Points are only plotted where a database match was found.

**Figure S3.D7. Energy above convex hull**

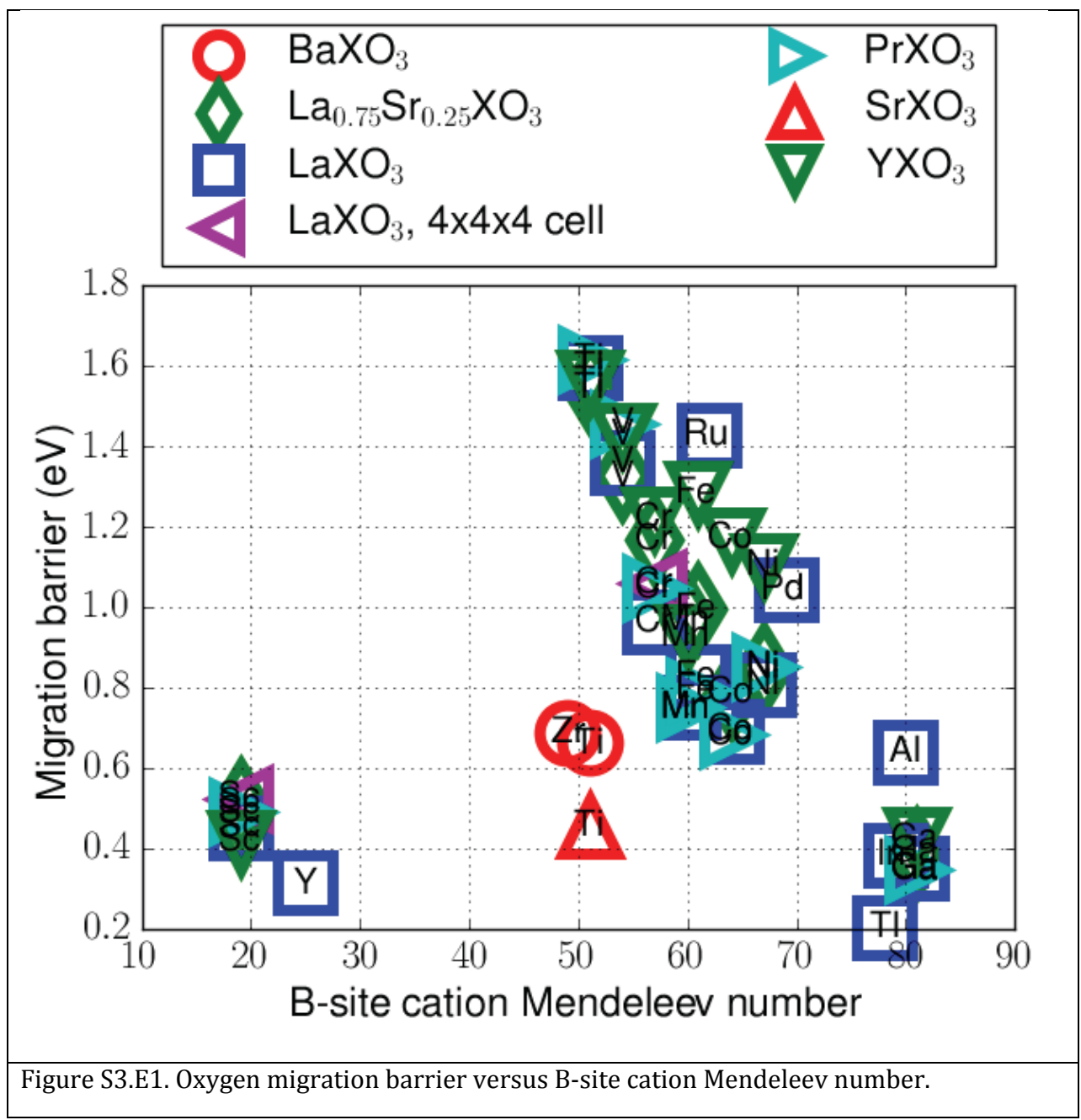

Figure S3.E1. Oxygen migration barrier versus B-site cation Mendeleev number.

Figure S3.E1. B-site cation Mendeleev number

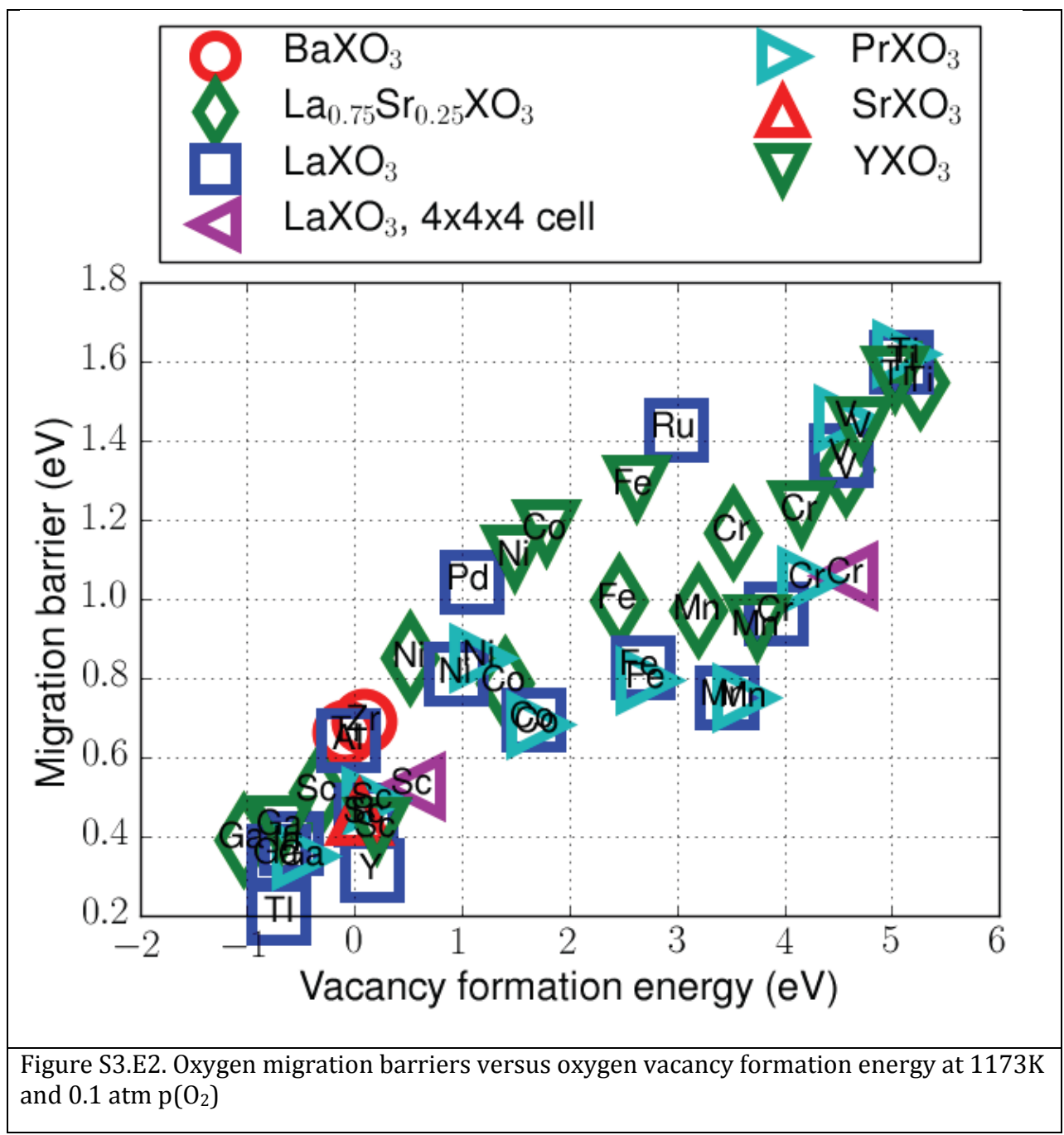

Figure S3.E2. Oxygen migration barriers versus oxygen vacancy formation energy at 1173K and 0.1 atm p($O_2$)

Figure S3.E2. Oxygen vacancy formation energy

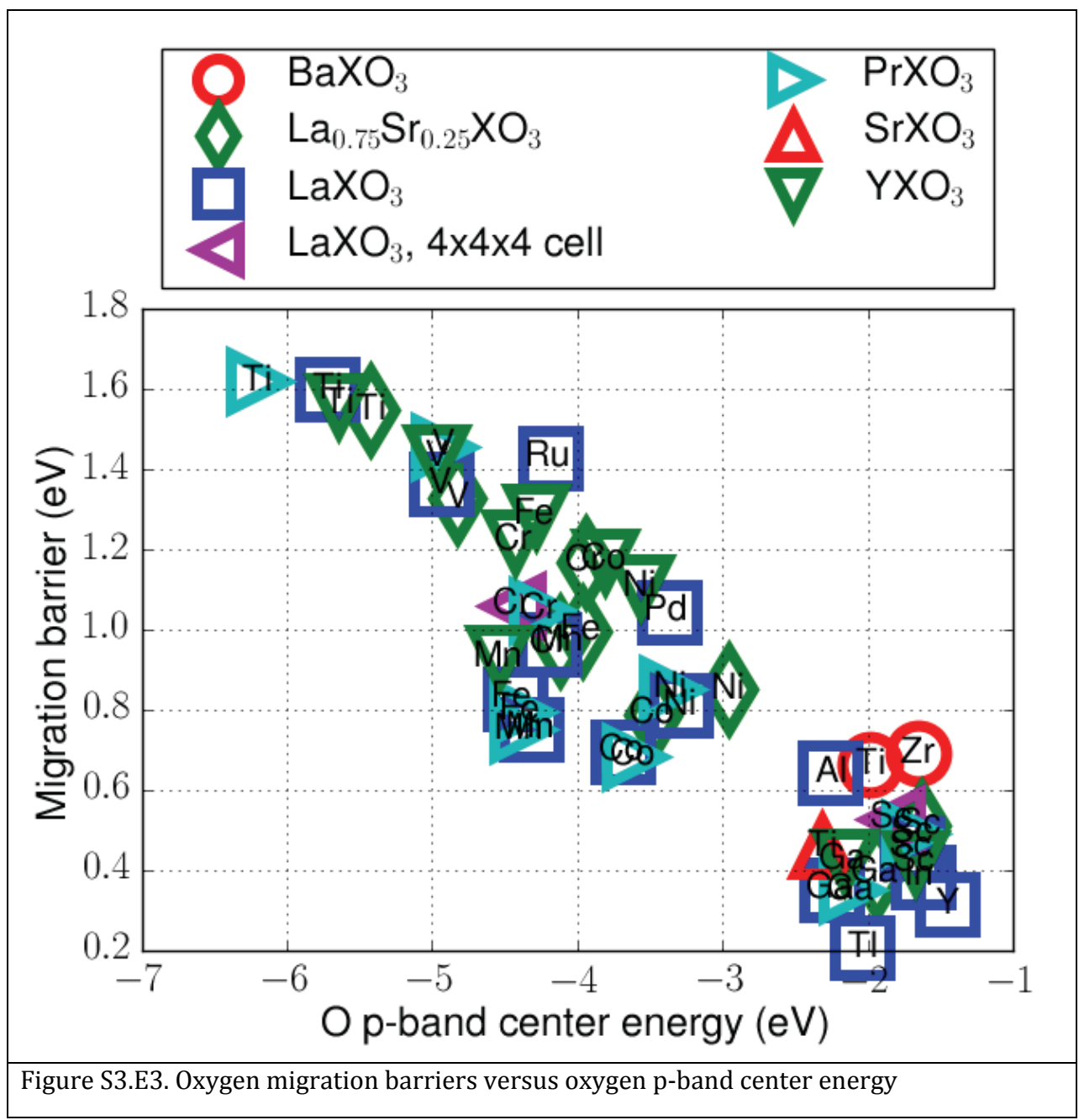

Figure S3.E3. Oxygen migration barriers versus oxygen p-band center energy

Figure S3.E3. Oxygen p-band center energy

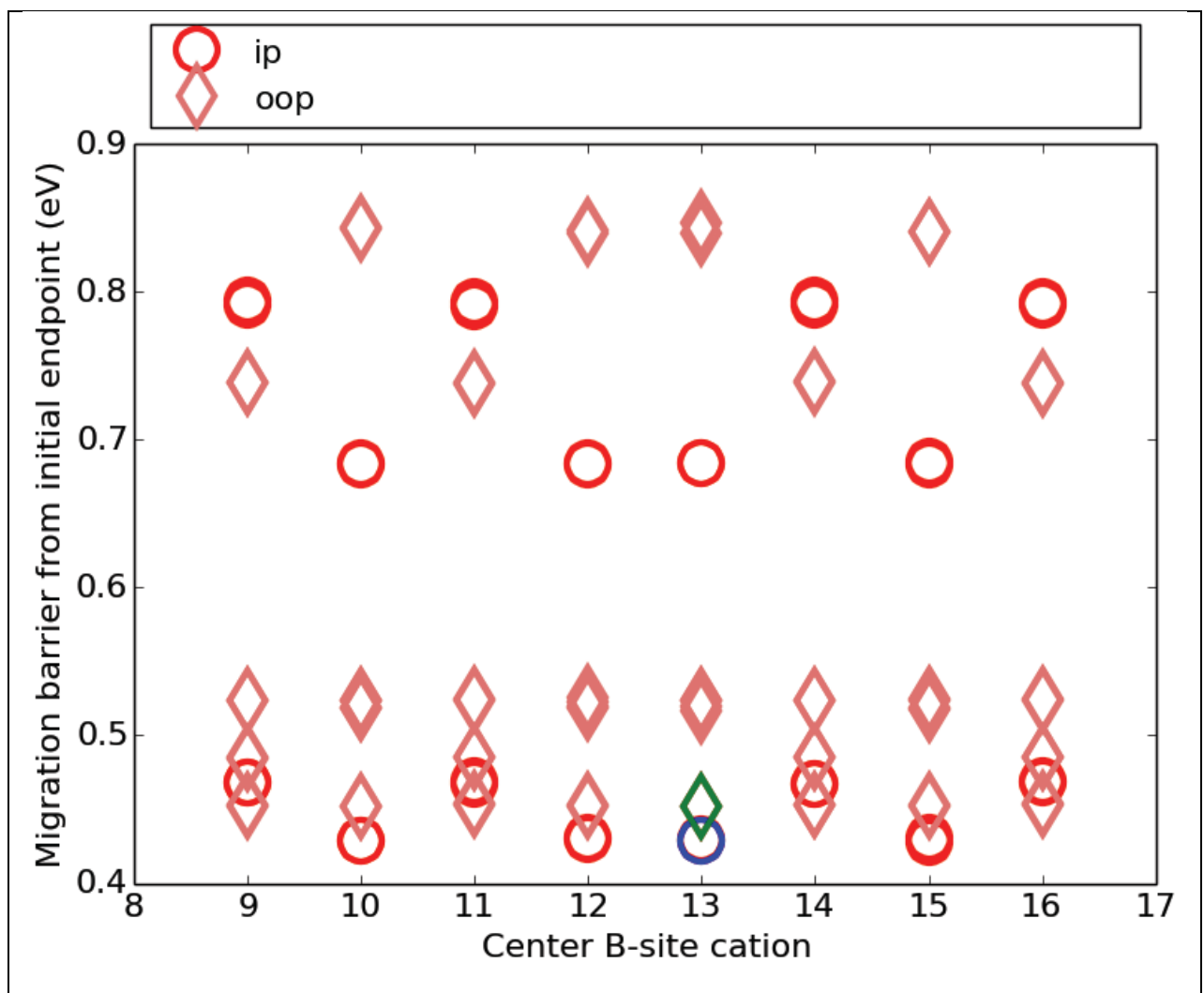

Figure S4.1. Barriers for 96 hops in LaScO$_3$, compensated. Some data points overlap. 'ip' and 'oop' designate hops that match in-plane and out-of-plane hops in the biaxial strain study of Ref.[3], and the in-plane hop used for all systems and the out-of-plane hop used for all systems (described in Ref. [3] Section S8) are highlighted in blue and green, respectively

**Figure S4.1. Barriers for 96 hops in LaScO$_3$, compensated.**

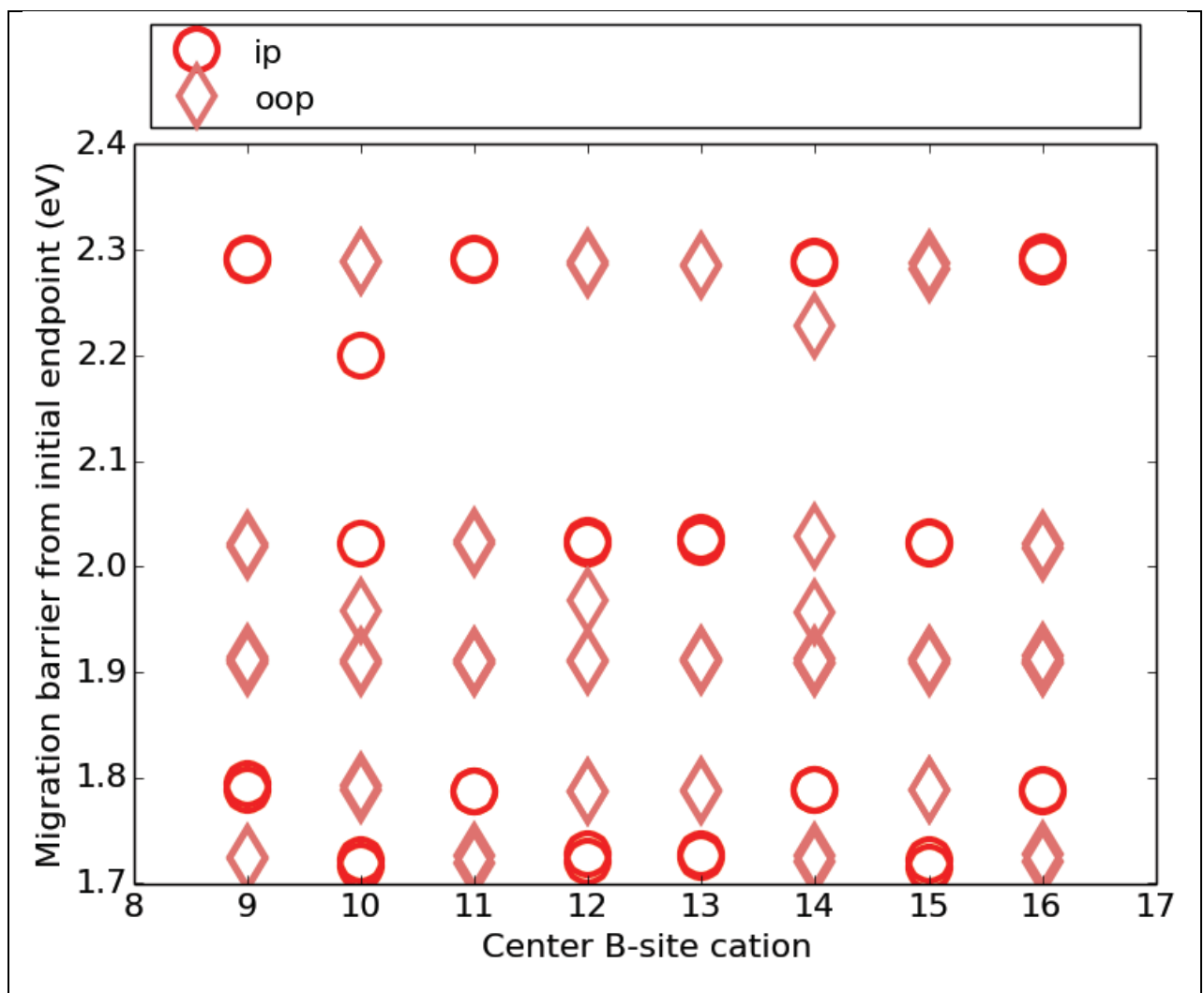

Figure S4.2. Barriers for 96 hops in LaScO$_3$, uncompensated. Some data points overlap. 'ip' and 'oop' designate hops that match in-plane and out-of-plane hops in the biaxial strain study of Ref.[3]

**Figure S4.2. Barriers for 96 hops in LaScO$_3$, uncompensated.**

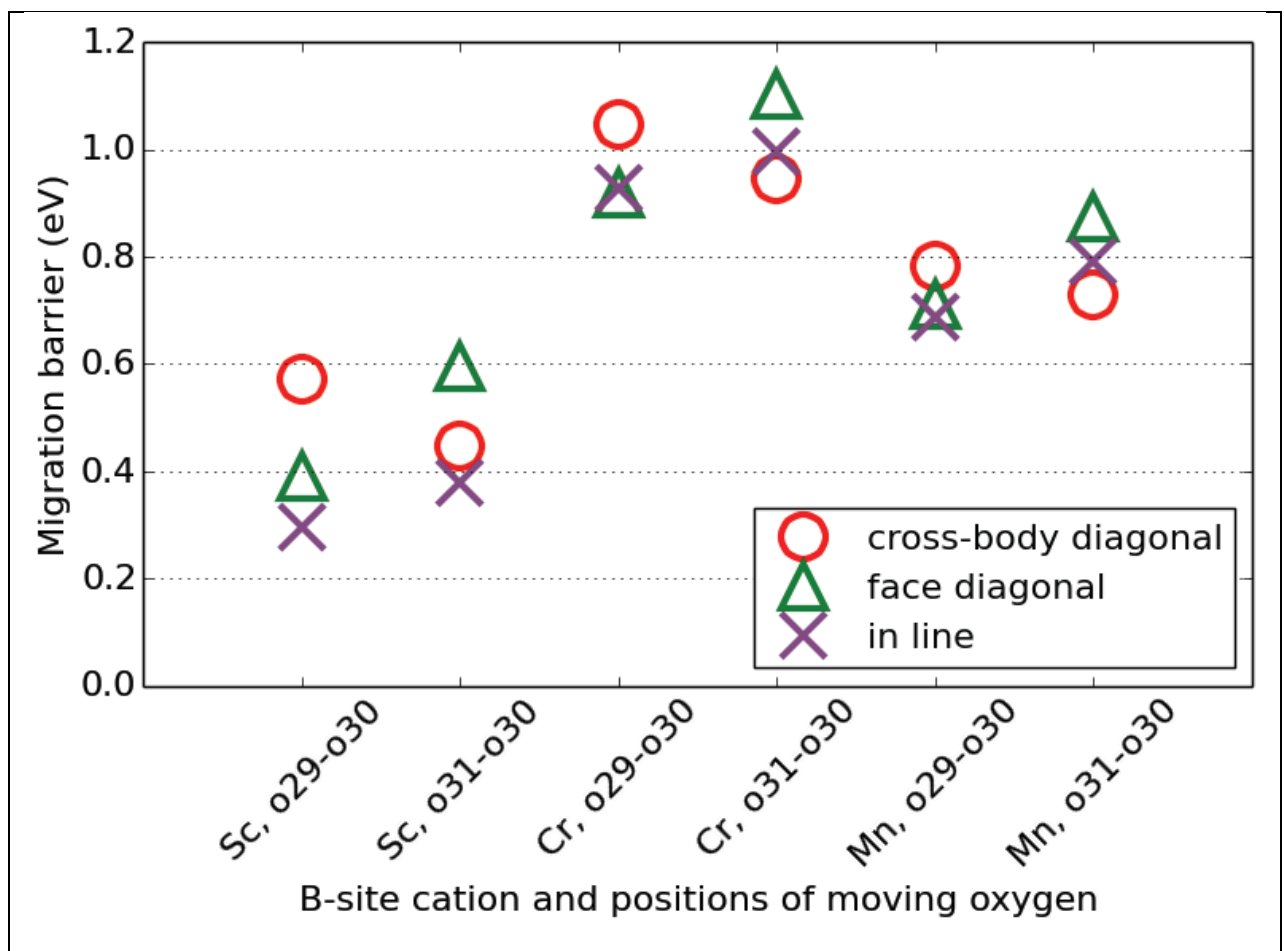

Figure S4.3. Effect of dopant positions on two migration barriers in each of three $La_{0.75}Sr_{0.25}XO_3$ systems. Dopant positions are from the ESI of Ref.[3] and are a1 and a8, a2 and a8, and a4 and a8 for cross-body diagonal, face diagonal, and in-line, respectively. The largest spread in migration barriers per system is 0.3 eV.

**Figure S4.3. Effect of dopant positions on migration barriers**

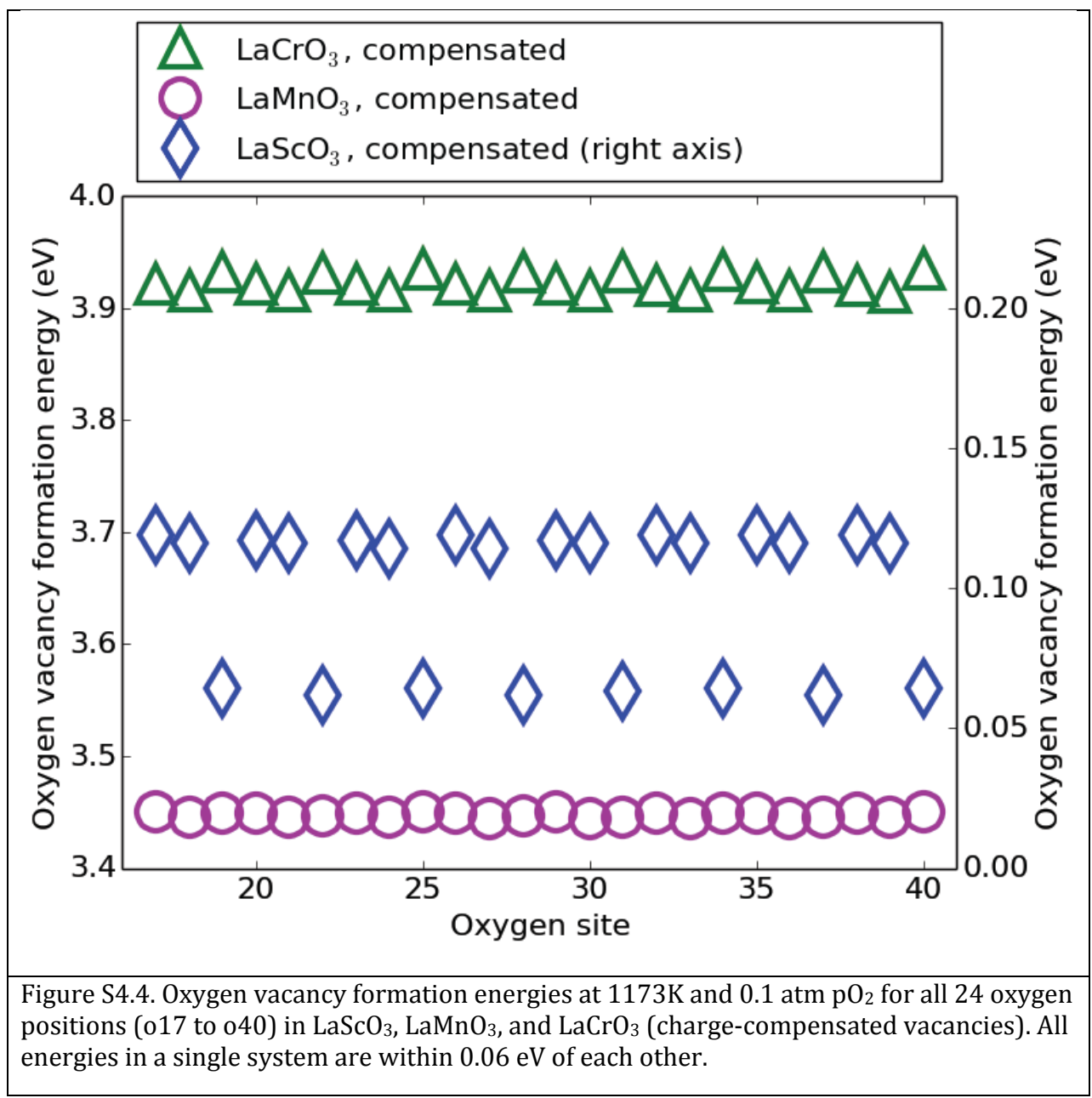

Figure S4.4. Oxygen vacancy formation energies at 1173K and 0.1 atm $pO_2$ for all 24 oxygen positions (o17 to o40) in $LaScO_3$, $LaMnO_3$, and $LaCrO_3$ (charge-compensated vacancies). All energies in a single system are within 0.06 eV of each other.

**Figure S4.4. All oxygen vacancy formation energies in three systems.**

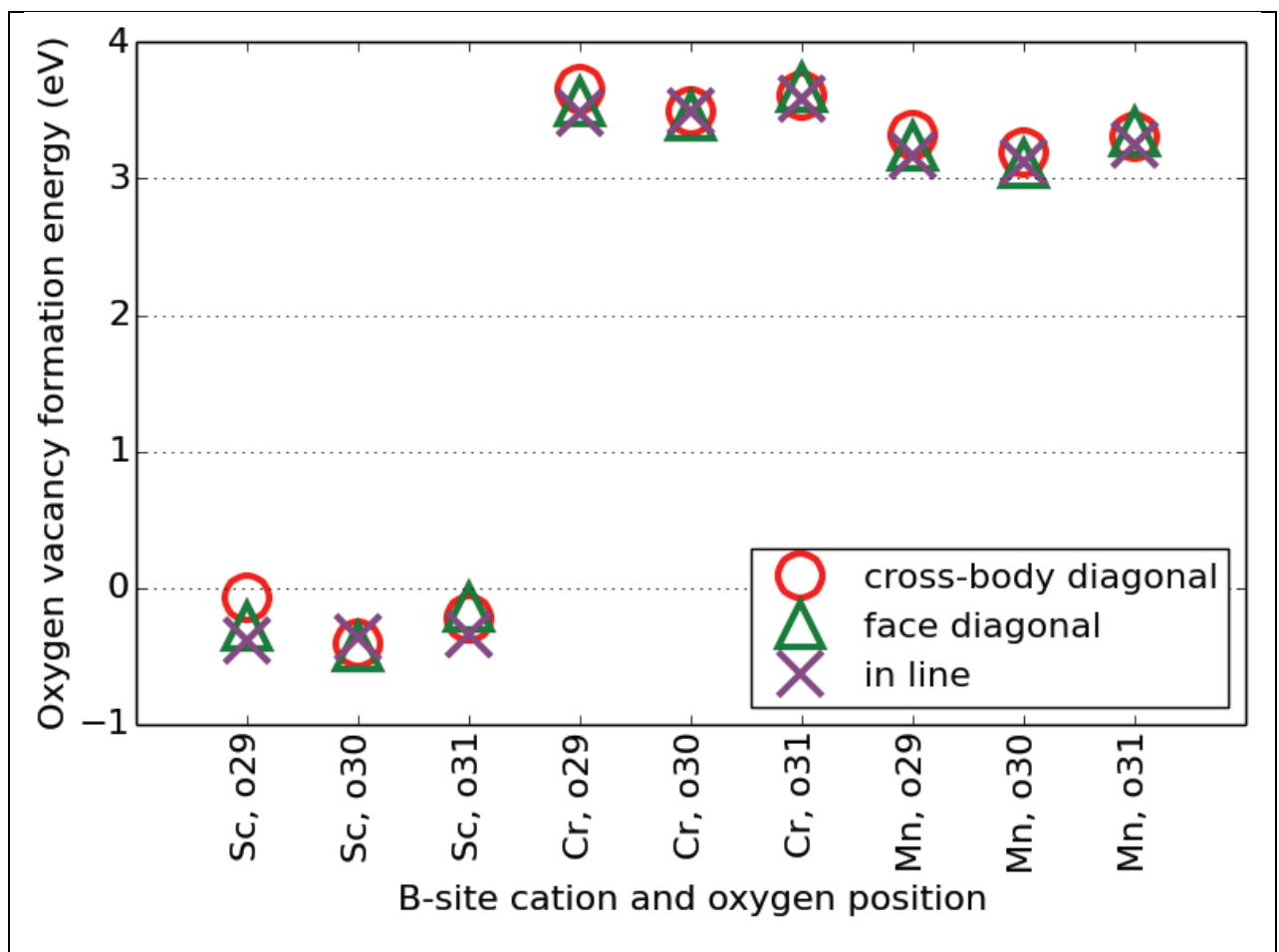

Figure S4.5. Effect of dopant positions on vacancy formation energy in $La_{0.75}Sr_{0.25}XO_3$. Dopant positions are described in Figure S4.3. The largest spread in vacancy formation energy per system over the three oxygen positions and three doping configurations is 0.4 eV.

Note that the negative vacancy formation energies for $La_{0.75}Sr_{0.25}ScO_3$ signify the propensity of the system to become $La_{0.75}Sr_{0.25}ScO_{3-\delta}$.

**Figure S4.5. Effect of dopant positions on vacancy formation energy**